\documentclass[aps,amsmath,twocolumn,amssymb,titlepage,reprint]{revtex4-1}
\usepackage[utf8]{inputenc}
\usepackage{amsmath}
\usepackage{amsfonts}
\usepackage{amssymb}
\usepackage{amsthm}
\usepackage{graphicx}
\usepackage{natbib}
\usepackage{float}
\usepackage{indentfirst}
\usepackage{booktabs}
\usepackage{multirow}
\usepackage[colorlinks = true,
            linkcolor = blue,
            urlcolor  = blue,
            citecolor = blue,
            anchorcolor = blue]{hyperref}
\usepackage{xcolor} 

\include{header}

\begin{document}
\title{Symmetry Analysis of Tensors in the Honeycomb Lattice of Edge-Sharing Octahedra}
\author{Franz G. Utermohlen}
\email[E-mail:]{utermohlen.1@osu.edu}
\author{Nandini Trivedi}
\email[E-mail:]{trivedi.15@osu.edu}
\affiliation{Department of Physics, The Ohio State University, Columbus, OH-43210, USA}

\begin{abstract}
We obtain the most general forms of rank-2 and rank-3 tensors allowed by the crystal symmetries of the honeycomb lattice of edge-sharing octahedra for crystals belonging to different crystallographic point groups, including the monoclinic point group $2/m$ and the trigonal (or rhombohedral) point group $\bar{3}$. Our results are relevant for two-dimensional materials, such as $\alpha$-RuCl$_3$, CrI$_3$, and the honeycomb iridates. We focus on the magnetic-field-dependent thermal conductivity tensor $\kappa_{ij}(\mathbf{H})$, which describes a system's longitudinal and thermal Hall responses, for the cases when the magnetic field is applied along high-symmetry directions, perpendicular to the plane and in the plane. We highlight some unexpected results, such as the equality of fully-longitudinal components to partially-transverse components in rank-3 tensors for systems with three-fold rotational symmetry, and make testable predictions for the thermal conductivity tensor.

\end{abstract}

\maketitle

\tableofcontents

\section{Introduction}
\begin{figure*}[ht!]
\definecolor{darkblue}{RGB}{86,73,157}
\definecolor{violet}{RGB}{174,35,157}
\definecolor{pink}{RGB}{234,146,224}
\definecolor{orange_symmetries}{RGB}{255,137,44}
\definecolor{red_bond}{RGB}{251,57,56}
\definecolor{green_bond}{RGB}{81,241,93}
\definecolor{blue_bond}{RGB}{0,80,238}
\begin{center}
\includegraphics[width=5.5in]{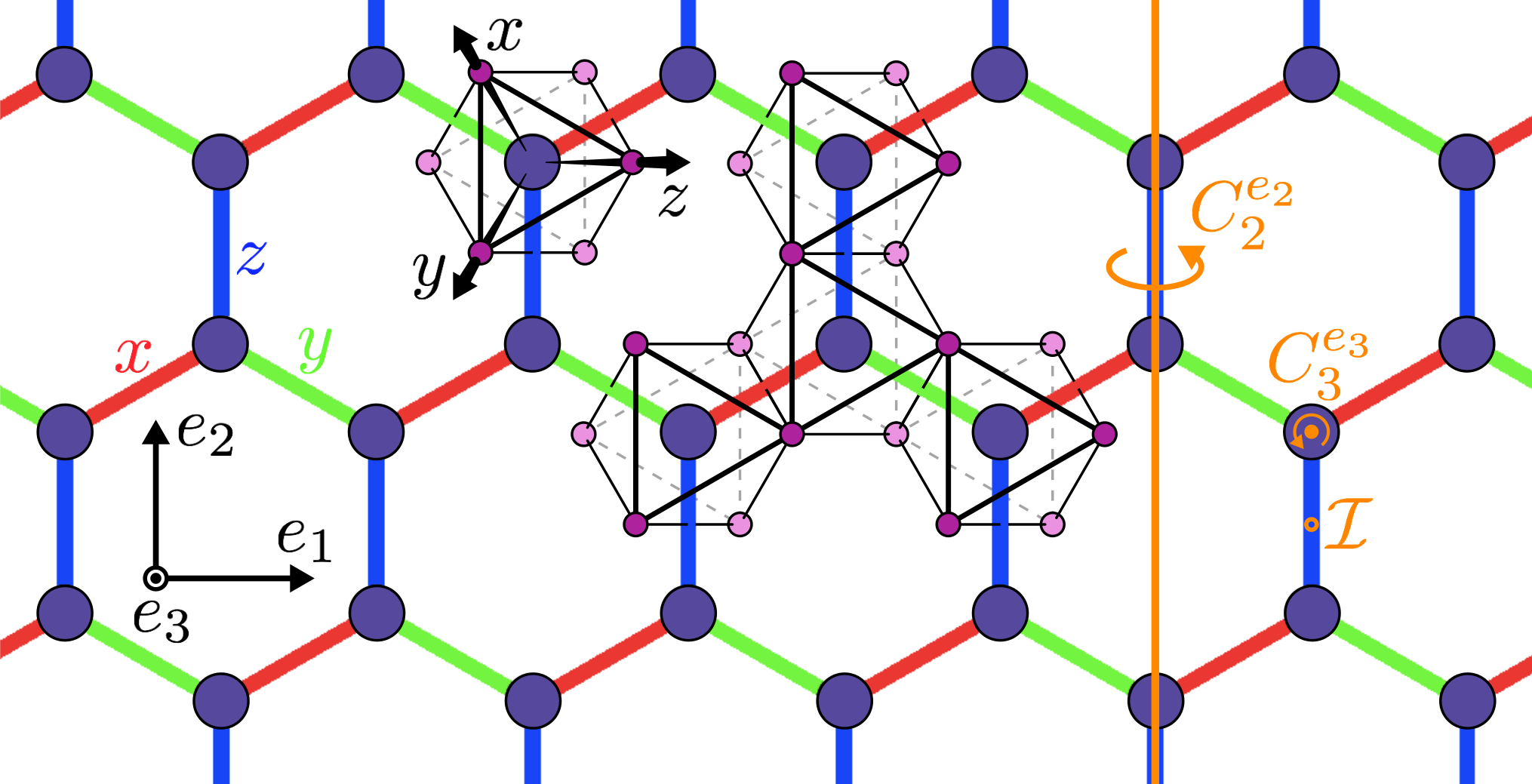}
\caption{Honeycomb lattice of edge-sharing octahedra and its possible point group generating symmetries. The \mbox{\textcolor{darkblue}{\textbf{dark blue}}} circles are magnetic metal ions, and the \textcolor{violet}{\textbf{violet}} and \textcolor{pink}{\textbf{pink}} circles surrounding them are ligands above and below the honeycomb plane, respectively, that form octahedra.
Each metal ion interacts with its three neighboring metal ions through superexchange mediated by their shared non-magnetic ligands. This interaction is generally bond-dependent, so we can label the three different types of bonds as $x$-, $y$-, and $z$-bonds (shown in \textcolor{red_bond}{\textbf{red}}, \textcolor{green_bond}{\textbf{green}}, and \textcolor{blue_bond}{\textbf{blue}}, respectively).
The three possible generating symmetries of the crystallographic point groups associated with this lattice are shown in \textcolor{orange_symmetries}{\textbf{orange}}.
$C_2^{e_2}$ is the two-fold ($180^\circ$) rotational symmetry with respect to each $z$-bond axis,
$C_3^{e_3}$ is the three-fold ($120^\circ$) rotational symmetry with respect to the out-of-plane axis through each site,
and $\mathcal{I}$ is the inversion symmetry with respect to each bond center.}
\label{fig:honeycomb_lattice_of_edge-sharing_octahedra}
\end{center}
\end{figure*}

Two-dimensional (2D) van der Waals crystals have been an active area of study ever since the recent discovery of 2D magnetism~\cite{Zhang2015,Huang2017,Gong2017,Burch2018,Lado2017,Huang2018,JiangLi2018,Wang2018,Liu2018,Chen2018,LeeUtermohlen2020,Chen2020,McCreary2020,Soriano2020},
quantum spin liquids (QSL)~\cite{Kitaev2006,Jackeli2009,Chaloupka2010,Rau2014,Banerjee2016,Banerjee2017,Baek2017,Banerjee2018,Takagi2019}, and
topological properties~\cite{KaneMele2005,HasanKane2010,LuVishwanath2012,Owerre2016,Kou2017,Pershoguba2018,McClarty2018,Chen2018}
in these materials.
In particular, the 2D van der Waals material \mbox{$\alpha$-RuCl$_3$} has attracted a great deal of attention because it is a close physical realization of the Kitaev honeycomb model~\cite{Plumb2014,Sears2015}, which is known to host a QSL phase, and as such it has been experimentally observed to have a QSL phase in the presence of an external magnetic field~\cite{Banerjee2016,Banerjee2017,Baek2017,Banerjee2018}.
Recently, a half-quantized thermal Hall effect was observed in the field-induced QSL phase of \mbox{$\alpha$-RuCl$_3$}~\cite{KasaharaOhnishi2018,Yokoi2020} for the magnetic field applied along different directions.
Several theoretical works have also explored the effect of
a magnetic field along different directions
on the Kitaev QSL~\cite{Janssen2019,HickeyTrebst2019,Ronquillo2019,Hickey2021,Gohlke2018,Gordon2019,PatelTrivedi2019,NasuMotome2019,Pradhan2020,JiangWang2018}.
Motivated by these experiments, we perform a symmetry-based tensor analysis on the honeycomb lattice of edge-sharing octahedra~(Fig.~\ref{fig:honeycomb_lattice_of_edge-sharing_octahedra})
in order to understand the directional dependence of physical responses in \mbox{$\alpha$-RuCl$_3$}~\cite{Ozel2019} and other 2D van der Waals materials with similar crystal structure~\cite{McGuire2017,Winter2017,Trebst2017}, such as CrI$_3$~\cite{Huang2017,Lado2017,Huang2018,JiangLi2018,Wang2018,Liu2018,Chen2018,LeeUtermohlen2020,Chen2020,McCreary2020,Soriano2020} and the honeycomb iridates~\cite{OMalley2008,Choi2012,Singh2012,Rau2014,Chun2015,Winter2016}.

The physical behavior of a system can be described using response tensors, which contain information about how the system's properties respond to perturbations applied along different directions.
A common example of a tensor is the magnetic susceptibility tensor \mbox{$\chi_{ij} = (\partial M_i/\partial H_j)|_{\mathbf{H}=\mathbf{0}}$}, which describes how the $i$ component of the system's magnetization $\mathbf{M}$ changes when we apply a weak magnetic field $\mathbf{H}$ along the $j$ direction.

In this work we use the symmetries of the honeycomb lattice of edge-sharing octahedra (see Fig.~\ref{fig:honeycomb_lattice_of_edge-sharing_octahedra}) to
obtain the most general forms of \mbox{rank-2} and \mbox{rank-3} response tensors allowed in such systems.
The driving principle in our analysis is that the crystal's physical properties obey the crystal symmetries, often referred to as Neumann's principle~\cite{Birss1964,Post1978,Shapiro2015,SorensenFisher2020}, and thus tensors describing its behavior remain invariant under the corresponding symmetry transformations.
This allows us to find the constraints imposed by each crystal symmetry on the tensor components.

We consider tensors describing systems with and without external fields (magnetic or electric) applied, which we will refer to as \mbox{\textit{field-dependent tensors}} and \mbox{\textit{zero-field tensors}}, respectively.
We also specifically examine the general form of the magnetic-field-dependent thermal conductivity tensor $\kappa_{ij}(\mathbf{H})$~\cite{AkgozSaunders1975_pt_I}, which describes a system's longitudinal and thermal Hall responses, as it is of current experimental interest~\cite{Hentrich2020,Yokoi2020,Hentrich2019,KasaharaOhnishi2018,KasaharaSugii2018,Hentrich2018}.

We note that there is a subtle but important distinction between a field-dependent tensor and a zero-field tensor that describes an experiment in which a small external field is used as a probe.
For example, the magnetic susceptibility tensor \mbox{$\chi_{ij} = (\partial M_i/\partial H_j)|_{\mathbf{H}=\mathbf{0}}$} is a zero-field tensor, \textit{not} a field-dependent tensor, even though its definition contains a magnetic field derivative.
This is because the magnetic field being applied here is infinitesimally small and therefore does not alter the system's ground state; it only serves to probe the properties of the system's zero-field ground state.
On the other hand, the field-dependent tensors describe the response of the finite-field ground state to an infinitesimal perturbation.
This is not just a conceptual distinction, but also a mathematical distinction: tensors by definition transform linearly with respect to the vector indices they are composed of, whereas field-dependent tensors in general do not transform linearly with the fields on which they are functionally dependent.

For ease of comparison to experiments, we work in the Cartesian coordinates $e_1 e_2 e_3$, where $e_1$ is a zigzag direction, $e_2$ is the armchair direction perpendicular to $e_1$, and $e_3$ is the direction perpendicular to the plane (see Fig.~\ref{fig:honeycomb_lattice_of_edge-sharing_octahedra}).
These coordinates are related to the octahedral coordinates $xyz$ through
\begin{equation}
\begin{aligned}
\hat{\mathbf{e}}_1 &\equiv \frac{1}{\sqrt{6}}( - \hat{\mathbf{x}} - \hat{\mathbf{y}} + 2\hat{\mathbf{z}}) \,, \\
\hat{\mathbf{e}}_2 &\equiv \frac{1}{\sqrt{2}}(\hat{\mathbf{x}} - \hat{\mathbf{y}}) \,, \\
\hat{\mathbf{e}}_3 &\equiv \frac{1}{\sqrt{3}}(\hat{\mathbf{x}} + \hat{\mathbf{y}} + \hat{\mathbf{z}}) \,.
\end{aligned}
\end{equation}

The paper is organized as follows:
\begin{itemize}
    \item Section~\ref{section:symmetries_and_point_groups}: we describe the symmetries possible in the honeycomb lattice of edge-sharing octahedra and list the crystallographic point groups generated by these symmetries.
    \item Section~\ref{section:zero-field_tensors}: we describe the constraints placed by these symmetries on the general forms of \mbox{rank-2} and \mbox{rank-3} for systems with no external fields, and make testable predictions for the magnetic field derivative of the thermal conductivity tensor, \mbox{$(\partial \kappa_{ij}/\partial H_k)|_{\mathbf{H}=\mathbf{0}}$}.
    \item Section~\ref{section:field-dependent_tensors}: we describe the types of symmetry constraints placed on tensors for systems with external fields, as well as the symmetry constraints on the magnetic-field-dependent thermal conductivity tensor $\kappa_{ij}(\mathbf{H})$.
    \item Section~\ref{section:summary_of_predictions_for_experiments}: we summarize the main predictions of this paper.
    \item Section~\ref{section:outlook}: we discuss potential uses and future directions for these results.
\end{itemize}

\section{Symmetries and Point Groups}
\label{section:symmetries_and_point_groups}
The macroscopic properties of a crystal depend only on its point group symmetries (i.e., rotations, reflections, and inversions), and not on its translational or space group symmetries~\cite{Birss1964}.
We therefore only have to consider these symmetries in our analysis.
For simplicity, in our analysis we only work with crystallographic point groups, and not with magnetic point groups, although we do consider the effect of magnetization on the system's symmetries.
All of the point group symmetries of the ideal honeycomb lattice of edge-sharing octahedra can be obtained from combinations of just three \mbox{\textit{generating symmetries}}, so it will only be necessary to consider the constraints placed on response tensors by these three symmetries.

The three generating symmetries in the ideal honeycomb lattice of edge-sharing octahedra are \mbox{$\{C_2^{e_2},C_3^{e_3},\mathcal{I}\}$}~\footnote{The choice of which three point group symmetries we use as the generating symmetries is not unique; they just have to be linearly independent. For example, we could have used a mirror symmetry instead of the inversion symmetry.}, which are described in more detail in the three subsections below.
We will also consider the cases where some or all of these generating symmetries are broken, as is often the case in materials.
A list of the crystallographic point groups formed by all of the subsets of these generating symmetries and examples of materials that belong to these point groups is given in \mbox{Table~\ref{tab:point_groups}}.

\begin{table*}[ht!]
\begin{center}
\begin{tabular}{|c|c|c|l|l|}
\hline
\multirow{3}{*}{Crystal System}
&
\multicolumn{2}{c|}{Point Group}
&
\multicolumn{1}{c|}{\multirow{3}{*}{Generating Symmetries}}
&
\multicolumn{1}{c|}{\multirow{3}{*}{Examples of Materials}}
\\ \cline{2-3}
&
\begin{tabular}{c}\vspace{-2pt}Hermann--Mauguin\\Symbol\vspace{1pt}\end{tabular}
&
\begin{tabular}{c}\vspace{-2pt}Schoenflies\\Symbol\vspace{1pt}\end{tabular}
&
&
\\ \hline
\multirow{2.2}{*}{Triclinic} & $1$ & $C_1$ & \hspace{20pt} $E$ & --- \\[1pt] \cline{2-5} 
 & $\bar{1}$ & $C_i$ ($S_2$) & \hspace{20pt} $\mathcal{I}$ & --- \\[1pt] \hline
\multirow{4.7}{*}{Monoclinic} & $2$ & $C_2$ & \hspace{20pt} $C_2^{e_2}$ & --- \\[1pt] \cline{2-5} 
 & $2/m$ & $C_{2h}$ & \hspace{20pt} $C_2^{e_2}$, $\mathcal{I}$ & \begin{tabular}{@{}l@{}}$\alpha$-RuCl$_3$~\cite{Johnson2015}, CrI$_3$~\cite{McGuire2015}, CrCl$_3$~\cite{McGuireClark2017}, Na$_2$IrO$_3$~\cite{Choi2012}, \\ $\alpha$-Li$_2$IrO$_3$~\cite{OMalley2008}, FePS$_3$~\cite{Lee2016}, IrCl$_3$, IrBr$_3$, IrI$_3$, \\ AlCl$_3$, MoCl$_3$, RhCl$_3$, RhBr$_3$, RhI$_3$, TcCl$_3$~\cite{McGuire2017}\end{tabular} \\[8pt] \hline
\multirow{7.4}{*}{Trigonal} & $3$ & $C_3$ & \hspace{20pt} $C_3^{e_3}$ & --- \\[1pt] \cline{2-5} 
 & $\bar{3}$ & $C_{3i}$ ($S_6$) & \hspace{20pt} $C_3^{e_3}$, $\mathcal{I}$ & \begin{tabular}{@{}l@{}}CrI$_3$~\cite{McGuire2015,Ubrig2020}, CrCl$_3$~\cite{McGuireClark2017}, CrBr$_3$, VCl$_3$, VBr$_3$~\cite{McGuire2017}, \\ VI$_3$~\cite{Kong2019,Dolezal2019}, BiI$_3$, FeCl$_3$, TiCl$_3$, TiBr$_3$, Ti$_3$O~\cite{McGuire2017}, \\ Cr$_2$Ge$_2$Te$_6$~\cite{Carteaux1995}, MnPSe$_3$, FePSe$_3$~\cite{Wiedenmann1981}\end{tabular} \\[8pt] \cline{2-5} 
 & $32$ & $D_3$ & \hspace{20pt} $C_2^{e_2}$, $C_3^{e_3}$ & \begin{tabular}{@{}l@{}}FeCl$_3$~\cite{McGuire2017}\end{tabular} \\[2pt] \cline{2-5} 
 & $\bar{3}m$ & $D_{3d}$ & \hspace{20pt} $C_2^{e_2}$, $C_3^{e_3}$, $\mathcal{I}$ & --- \\[1pt] \hline
\end{tabular}
\caption{Generating symmetries and materials associated with the eight possible crystallographic point groups for the honeycomb lattices of edge-sharing octahedra.
A given point group has rotation, reflection, and inversion symmetries that can be obtained from combinations of its generating symmetries.
The symmetries $C_2^{e_2}$, $C_3^{e_3}$, $\mathcal{I}$ are described in Fig.~\ref{fig:honeycomb_lattice_of_edge-sharing_octahedra}, and the symmetry $E$ simply corresponds to the identity operation, which leaves the system unchanged (i.e., no rotation, reflection, or inversion).
Trigonal crystals are sometimes referred to as rhombohedral, especially when they belong to a rhombohedral space group, such as $R\bar{3}$.
Some materials are listed under more than one point group because can have different crystal structures depending on their temperature and sample thickness.}
\label{tab:point_groups}
\end{center}
\end{table*}

\subsection{Two-Fold Rotational Symmetry (\texorpdfstring{$C_2^{e_2}$}{C2e2})}
The honeycomb lattice shown in Fig.~\ref{fig:honeycomb_lattice_of_edge-sharing_octahedra} can have two-fold rotational symmetry
(i.e., $180^\circ$ rotational symmetry)
with respect to the armchair axis $e_2$ passing through each $z$-bond
($C_2^{e_2}$).
This symmetry transformation corresponds to a rotation by $180^\circ$ with respect to the armchair axis $e_2$ and is described by the coordinate rotation matrix
\begin{equation}
\mathbf{C}_2^{e_2} =
\begin{pmatrix}
\cos(180^\circ) & 0 & \sin(180^\circ) \\
0 & 1 & 0 \\
-\sin(180^\circ) & 0 & \cos(180^\circ) \\
\end{pmatrix}
=
\begin{pmatrix}
-1 & 0 & 0 \\
0 & 1 & 0 \\
0 & 0 & -1
\end{pmatrix} \,,
\end{equation}
which effectively reverses a vector's $e_1$ and $e_3$ components:
\begin{equation}
\begin{pmatrix}
v_{e_1} \\
v_{e_2} \\
v_{e_3}
\end{pmatrix} \xrightarrow{C_2^{e_2}}
\mathbf{C}_2^{e_2}
\begin{pmatrix}
v_{e_1} \\
v_{e_2} \\
v_{e_3}
\end{pmatrix}
=
\begin{pmatrix}
-v_{e_1} \\
v_{e_2} \\
-v_{e_3}
\end{pmatrix} \,.
\end{equation}

We note that in crystals with $C_2^{e_2}$ symmetry (i.e., belonging to the monoclinic point groups $2$ or $2/m$, or to the trigonal point groups $32$, or $\bar{3}m$), the system's $C_2^{e_2}$ symmetry can still be broken if it is magnetized along an axis that does not have $C_2$ symmetry.

\subsection{Three-Fold Rotational Symmetry (\texorpdfstring{$C_3^{e_3}$}{C3e3})}
This lattice can also have three-fold rotational symmetry
(i.e., $120^\circ$ rotational symmetry)
with respect to the out-of-plane axis passing through each site
($C_3^{e_3}$).
This symmetry transformation is described by the coordinate rotation matrix
\begin{equation}
\mathbf{C}_3^{e_3} =
\begin{pmatrix}
\cos(120^\circ) & \sin(120^\circ) & 0 \\
-\sin(120^\circ) & \cos(120^\circ) & 0 \\
0 & 0 & 1
\end{pmatrix}
=
\begin{pmatrix}
-\frac{1}{2} & \frac{\sqrt{3}}{2} & 0 \\[2pt]
-\frac{\sqrt{3}}{2} & -\frac{1}{2} & 0 \\[1pt]
0 & 0 & 1
\end{pmatrix}
\,,
\end{equation}
which mixes a vector's in-plane components:
\begin{equation}
\begin{pmatrix}
v_{e_1} \\
v_{e_2} \\
v_{e_3}
\end{pmatrix} \xrightarrow{C_3^{e_3}}
\mathbf{C}_3^{e_3}
\begin{pmatrix}
v_{e_1} \\
v_{e_2} \\
v_{e_3}
\end{pmatrix}
=
\begin{pmatrix}
-\frac{1}{2}v_{e_1} + \frac{\sqrt{3}}{2}v_{e_2} \\[2pt]
-\frac{\sqrt{3}}{2}v_{e_1} - \frac{1}{2}v_{e_2} \\
v_{e_3}
\end{pmatrix} \,.
\end{equation}

We note that in crystals with $C_3^{e_3}$ symmetry (i.e., belonging to the trigonal point groups $3$, $\bar{3}$, $32$ and $\bar{3}m$), the system's $C_3^{e_3}$ symmetry can still be broken if it is magnetized along an axis other than the out-of-plane axis.

\begin{figure}[ht!]
\definecolor{orange_symmetries}{RGB}{255,137,44}
\begin{center}
\includegraphics[width=3.3in]{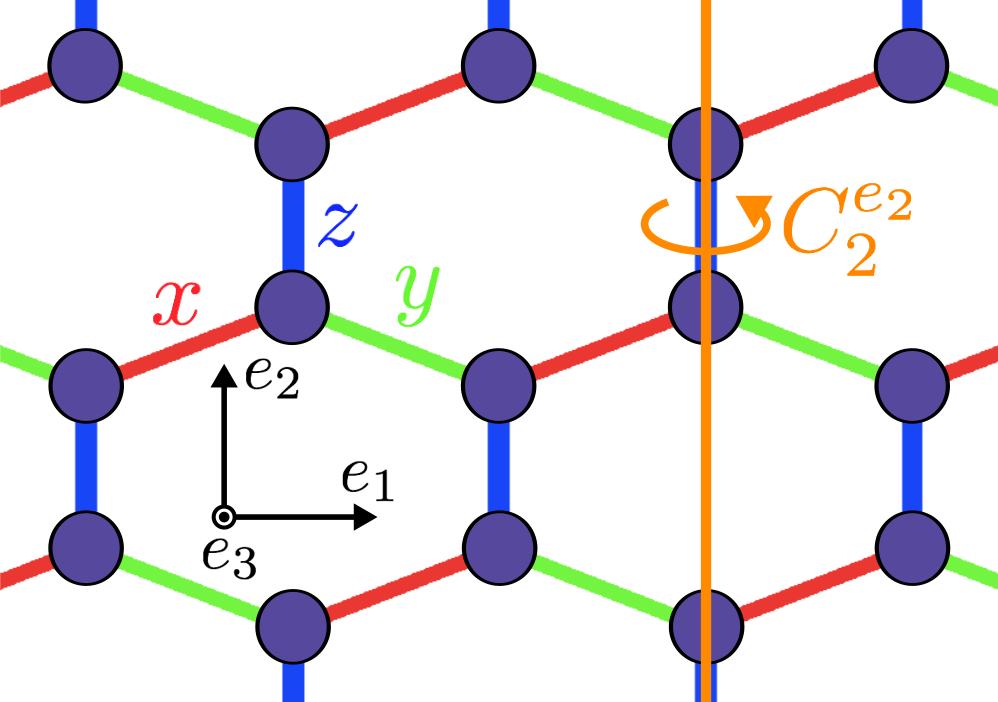}
\caption{Honeycomb lattice for a monoclinic crystal (point groups $2$ or $2/m$). This lattice lacks three-fold ($120^\circ$) rotational symmetry because $z$-bonds do not have the same length as $x$- and $y$-bonds, and it has two-fold ($180^\circ$) rotational symmetry because $x$- and $y$-bonds have the same length.
The high-symmetry armchair axis (shown in \textcolor{orange_symmetries}{\textbf{orange}}) along $z$-bonds therefore has two-fold rotational symmetry, whereas the other two armchair axes along $x$- and $y$-bonds lack this symmetry.}
\label{fig:honeycomb_lattice_of_edge-sharing_octahedra-squished}
\end{center}
\end{figure}

We now clarify a possible point of confusion in our definition of the in-plane axes $e_1$ and $e_2$.
For systems with $C_2^{e_2}$ symmetry but without $C_3^{e_3}$ symmetry (i.e., belonging to the monoclinic point groups $2$ and $2/m$), there are two different types of armchair axes: the unique armchair axis that has $C_2$ symmetry, and the other two equivalent armchair axes that do not have this symmetry (see Fig.~\ref{fig:honeycomb_lattice_of_edge-sharing_octahedra-squished}).
For these systems, we define $e_2$ as this unique high-symmetry armchair axis,
and similarly we define $z$-bonds as the bonds oriented along this axis.
For systems with $C_3^{e_3}$ symmetry (i.e., belonging to the trigonal point groups $3$, $\bar{3}$, $32$ and $\bar{3}m$), the three armchair axes are equivalent, so we arbitrarily define $e_2$ as any one of these axes.
Finally, for systems without $C_2^{e_2}$ or $C_3^{e_3}$ symmetry (i.e., belonging to the triclinic point groups $1$ and $\bar{1}$), the three armchair axes are all different, so we again arbitrarily define $e_2$ as any one of these axes.
In all of these cases, we define $e_1$ as the zigzag axis perpendicular to the $e_2$-axis.

\subsection{Inversion Symmetry (\texorpdfstring{$\mathcal{I}$}{I})}
Finally, this system can also have
bond-centered inversion symmetry ($\mathcal{I}$).
Under inversion, vectors transform as
\begin{equation}
\mathbf{v}
\xrightarrow{\mathcal{I}}
\lambda_\mathcal{I} \mathbf{v} \,, \label{eqn:inversion-eigenvalue_transformation_equation}
\end{equation}
where the eigenvalue $\lambda_\mathcal{I}=\pm 1$ depends on the particular vector $\mathbf{v}$.
Vectors that are odd under inversion \mbox{($\lambda_\mathcal{I}=-1$)} are called \textit{polar vectors} and include quantities such as electric field, electric current, temperature gradient, heat current, spin current, and momentum,
whereas vectors that are even under inversion \mbox{($\lambda_\mathcal{I}=+1$)} are called \textit{axial vectors} (or \textit{pseudovectors}) and include quantities like magnetic field, magnetization, and spin.

A more common way of describing how vectors transform under inversion is using the coordinate inversion matrix
\begin{equation}
\boldsymbol{\mathcal{I}} =
\begin{pmatrix}
-1 & 0 & 0 \\
0 & -1 & 0 \\
0 & 0 & -1
\end{pmatrix}
\end{equation}
and using the transformation rules
\begin{equation}
\begin{aligned}
\mathbf{v} &\xrightarrow{\mathcal{I}} \boldsymbol{\mathcal{I}}\mathbf{v} = -\mathbf{v} \quad &&\text{(polar vector)} \,, \\
\mathbf{v} &\xrightarrow{\mathcal{I}} |\boldsymbol{\mathcal{I}}|\boldsymbol{\mathcal{I}}\mathbf{v} = \mathbf{v} \quad &&\hspace{1pt}\text{(axial vector)} \,,
\end{aligned}
\end{equation}
where
\mbox{$|\boldsymbol{\mathcal{I}}|=-1$}
is the determinant of the transformation matrix $\boldsymbol{\mathcal{I}}$.
This formulation is useful because it allows us to generalize the transformation rules for polar and axial vectors under any orthogonal transformation matrix $\mathbf{R}$ as
\begin{equation}
\begin{aligned}
\mathbf{v} &\xrightarrow{R} \mathbf{R}\mathbf{v} \quad &&\text{(polar vector)} \,, \\
\mathbf{v} &\xrightarrow{R} |\mathbf{R}|\mathbf{R}\mathbf{v} \quad &&\hspace{1pt}\text{(axial vector)} \,,
\end{aligned}
\label{eqn:polar_and_axial_vector_transformation_rule}
\end{equation}
where transformation matrices with
\mbox{$|\mathbf{R}|=+1$} describe rotations,
whereas those with
\mbox{$|\mathbf{R}|=-1$} describe improper rotations (i.e., the combination of a rotation and an inversion).

\section{Zero-Field Tensors}
\label{section:zero-field_tensors}
In this section we describe the general forms of tensors allowed by the symmetries described earlier for systems with no external magnetic or electric fields, and as an example we discuss and make testable predictions for the magnetic field derivative of the thermal conductivity tensor, \mbox{$(\partial \kappa_{ij}/\partial H_k)|_{\mathbf{H}=\mathbf{0}}$}.

\begin{table*}[ht!]
\begin{center}
\begin{tabular}{|c|c|c|c|}
\cline{1-4}
\multicolumn{1}{|c|}{\begin{tabular}{c}\vspace{-2pt}Generating\\Symmetry\vspace{1pt}\end{tabular}}
&
\begin{tabular}{c}\vspace{-2pt}Rank-2 Tensors\\(no external field)\vspace{1pt}\end{tabular}
&
\begin{tabular}{c}\vspace{-2pt}Rank-3 Tensors\\(no external field)\vspace{1pt}\end{tabular}
&
\begin{tabular}{c}\vspace{-2pt}Rank-$r$ Tensors\\(no external field)\vspace{1pt}\end{tabular}
\\ \hline
$C_2^{e_2}$
&
\,$T_{e_1 e_2}=T_{e_2 e_1}=T_{e_2 e_3}=T_{e_3 e_2}=0$\,
&
\begin{tabular}{@{}c@{}}
$T_{e_1 e_1 e_1} = T_{e_3 e_3 e_3} = 0$ \\
$T_{e_1 e_1 e_3} = T_{e_1 e_3 e_1} = T_{e_3 e_1 e_1} = 0$ \\
$T_{e_3 e_3 e_1} = T_{e_3 e_1 e_3} = T_{e_1 e_3 e_3} = 0$ \\
$T_{e_1 e_2 e_2} = T_{e_2 e_1 e_2} = T_{e_2 e_2 e_1} = 0$ \\
\,$T_{e_3 e_2 e_2} = T_{e_2 e_3 e_2} = T_{e_2 e_2 e_3} = 0$\,
\\ \vspace{-10pt} \\
\end{tabular}
&
\begin{tabular}{@{}c@{}}
$T_{i_1 i_2 \hdots i_r} = 0$ \\
if $N_{e_1} + N_{e_3}$ is odd \\ \vspace{-5pt} \\
($N_\alpha = $ \# of indices equal to $\alpha$)
\end{tabular}
\\ \hline
$C_3^{e_3}$
&
\begin{tabular}{@{}c@{}}
$T_{e_1 e_1}=T_{e_2 e_2}$ \\ $T_{e_1 e_2}=-T_{e_2 e_1}$ \\ $T_{e_1 e_3}=T_{e_3 e_1}=T_{e_2 e_3}=T_{e_3 e_2}=0$
\end{tabular}
&
\begin{tabular}{@{}c@{}}
$T_{e_1 e_1 e_1} = -T_{e_1 e_2 e_2} = -T_{e_2 e_1 e_2} = -T_{e_2 e_2 e_1}$ \\
$T_{e_2 e_2 e_2} = -T_{e_2 e_1 e_1} = -T_{e_1 e_2 e_1} = -T_{e_1 e_1 e_2}$ \\
$T_{e_1 e_1 e_3} = T_{e_2 e_2 e_3}$ \\
$T_{e_1 e_3 e_1} = T_{e_2 e_3 e_2}$ \\
$T_{e_3 e_1 e_1} = T_{e_3 e_2 e_2}$ \\
$T_{e_1 e_2 e_3} = -T_{e_2 e_1 e_3}$ \\
$T_{e_1 e_3 e_2} = -T_{e_2 e_3 e_1}$ \\
$T_{e_3 e_2 e_1} = -T_{e_3 e_1 e_2}$ \\
$T_{e_1 e_3 e_3} = T_{e_3 e_1 e_3} = T_{e_3 e_3 e_1} = 0$ \\
$T_{e_2 e_3 e_3} = T_{e_3 e_2 e_3} = T_{e_3 e_3 e_2} = 0$
\\ \vspace{-10pt} \\
\end{tabular}
&
No simple constraint
\\ \hline
$\mathcal{I}$
&
\begin{tabular}{@{}c@{}}
$T_{ij} = 0$ \\
if exactly one of the indices $(i,j)$ \\
corresponds to a polar vector
\end{tabular}
&
\begin{tabular}{@{}c@{}}
$T_{ijk} = 0$ \\
if exactly one or three of the indices $(i,j,k)$ \\
correspond to polar vectors
\end{tabular}
&
\begin{tabular}{@{}c@{}}
$T_{i_1 i_2 \hdots i_r} = 0$ \\
if the number of \\
indices corresponding to \\
polar vectors is odd
\end{tabular}
\\ \hline
\end{tabular}
\caption{Constraints imposed by the point group generating symmetries on zero-field tensors.
These symmetries are described in Fig.~\ref{fig:honeycomb_lattice_of_edge-sharing_octahedra}.}
\label{tab:constraints_imposed_by_crystal_symmetries}
\end{center}
\end{table*}

\subsection{Rank-2 Tensors}
We can express a general \mbox{rank-2} tensor $T_{ij}$ for a system with no external field as a \mbox{$3 \times 3$} matrix in $e_1 e_2 e_3$ coordinates as
\begin{equation}
T_{ij}
= \begin{pmatrix}
T_{e_1 e_1} & T_{e_1 e_2} & T_{e_1 e_3} \\
T_{e_2 e_1} & T_{e_2 e_2} & T_{e_2 e_3} \\
T_{e_3 e_1} & T_{e_3 e_2} & T_{e_3 e_3}
\end{pmatrix} \,.
\label{eqn:general_rank-2_tensor_in_e1e2e3_coordinates}
\end{equation}
Under an orthogonal transformation matrix $\mathbf{R}$, \mbox{rank-2} tensors transform as~\footnote{We are using the Einstein summation convention for repeated indices throughout this paper.}
\begin{equation}
T_{ij} \xrightarrow{R} |\mathbf{R}|^{N_\text{axial}} R_{im}R_{jn}T_{mn} \,,
\end{equation}
where $N_\text{axial}$ is the number of indices in the tensor $T$ corresponding to axial vectors.
In matrix notation, this equation is
\begin{equation}
\mathbf{T} \xrightarrow{R} |\mathbf{R}|^{N_\text{axial}} \mathbf{R}\mathbf{T}\mathbf{R}^\mathrm{T} \,,
\end{equation}
where $\mathbf{R}^\mathrm{T}$ denotes the transpose of $\mathbf{R}$.

For example, under $C_2^{e_2}$, \mbox{rank-2} tensors transform as
\begin{equation}
\mathbf{T} \xrightarrow{C_2^{e_2}} \mathbf{T}' = \mathbf{C}_2^{e_2}\mathbf{T}(\mathbf{C}_2^{e_2})^\mathrm{T} \,,
\end{equation}
or explicitly,
\begin{equation}
\begin{pmatrix}
T'_{e_1 e_1} & T'_{e_1 e_2} & T'_{e_1 e_3} \\
T'_{e_2 e_1} & T'_{e_2 e_2} & T'_{e_2 e_3} \\
T'_{e_3 e_1} & T'_{e_3 e_2} & T'_{e_3 e_3}
\end{pmatrix}
=
\begin{pmatrix}
 T_{e_1 e_1} & -T_{e_1 e_2} &  T_{e_1 e_3} \\
-T_{e_2 e_1} &  T_{e_2 e_2} & -T_{e_2 e_3} \\
 T_{e_3 e_1} & -T_{e_3 e_2} &  T_{e_3 e_3}
\end{pmatrix} \,.
\end{equation}
Invariance under this transformation \mbox{($T_{ij}=T'_{ij}$)} imposes the constraints
\begin{equation}
T_{e_1 e_2} = T_{e_2 e_1} = T_{e_2 e_3} = T_{e_3 e_2} = 0 \,. \qquad (C_2^{e_2}) \label{eqn:rank-2_zero-field_tensor_constraints-C2e2_symmetry}
\end{equation}

Similarly, $C_3^{e_3}$ imposes the constraints
\begin{equation}
\begin{aligned}
T_{e_1 e_1} &= T_{e_2 e_2} \,, \\
T_{e_1 e_2} &= -T_{e_2 e_1} \,, \\
T_{e_1 e_3} = T_{e_3 e_1} &= T_{e_2 e_3} = T_{e_3 e_2} = 0 \,.
\end{aligned}
\qquad (C_3^{e_3})
\label{eqn:rank-2_zero-field_tensor_constraints-C3e3_symmetry}
\end{equation}
We note that for systems with $C_3^{e_3}$ symmetry, \mbox{rank-2} zero-field tensors have continuous rotational symmetry with respect to the axis perpendicular to the plane, since they are invariant upon rotating the orientation of the in-plane axes $e_1$ and $e_2$ to point along any two perpendicular directions inside the plane:
\begin{equation}
\begin{pmatrix}
\cos\theta & \sin\theta & 0 \\
-\sin\theta & \cos\theta & 0 \\
0 & 0 & 1
\end{pmatrix}
\mathbf{T}
\begin{pmatrix}
\cos\theta & \sin\theta & 0 \\
-\sin\theta & \cos\theta & 0 \\
0 & 0 & 1
\end{pmatrix}^\mathrm{T}
=
\mathbf{T} \,.
\end{equation}
In these systems, \mbox{rank-2} physical responses (such as the magnetic susceptibility \mbox{$\chi_{ij} = (\partial M_i/\partial H_j)|_{\mathbf{H}=\mathbf{0}}$}) therefore behave the same way along all in-plane directions~\cite{Andrade2020}, including low-symmetry directions.

Finally, inversion symmetry does not constrain the form of \mbox{rank-2} tensors, but it does require that either none or both of the tensor indices \mbox{($i,j$)} correspond to polar vectors, otherwise the tensor will equal zero.

The most general forms of \mbox{rank-2} zero-field tensors for systems belonging to the eight point groups generated by these three symmetries (see \mbox{Table~\ref{tab:point_groups}}) are given in \mbox{Table~\ref{tab:general_forms_of_tensors}}.

\subsection{Rank-3 Tensors}
\label{subsection:rank-3_zero-field_tensors}
Higher-rank tensors, such as \mbox{rank-3} tensors, can arise in a \textit{multilinear} manner as a linear response to multiple perturbations, such as the bilinear response of the magnetization to a thermal gradient and an applied magnetic field. In addition, higher-rank tensors are necessary to describe higher-order or \textit{nonlinear} responses to a perturbation.

We can express a general \mbox{rank-3} tensor $T_{ijk}$ for a system with no external field as a set of three \mbox{$3 \times 3$} matrices in $e_1 e_2 e_3$ coordinates as
\begin{equation}
\begin{aligned}
T_{e_1 jk} &= \begin{pmatrix}
T_{e_1 e_1 e_1} & T_{e_1 e_1 e_2} & T_{e_1 e_1 e_3} \\
T_{e_1 e_2 e_1} & T_{e_1 e_2 e_2} & T_{e_1 e_2 e_3} \\
T_{e_1 e_3 e_1} & T_{e_1 e_3 e_2} & T_{e_1 e_3 e_3}
\end{pmatrix} \,, \\
T_{e_2 jk} &= \begin{pmatrix}
T_{e_2 e_1 e_1} & T_{e_2 e_1 e_2} & T_{e_2 e_1 e_3} \\
T_{e_2 e_2 e_1} & T_{e_2 e_2 e_2} & T_{e_2 e_2 e_3} \\
T_{e_2 e_3 e_1} & T_{e_2 e_3 e_2} & T_{e_2 e_3 e_3}
\end{pmatrix} \,, \\
T_{e_3 jk} &= \begin{pmatrix}
T_{e_3 e_1 e_1} & T_{e_3 e_1 e_2} & T_{e_3 e_1 e_3} \\
T_{e_3 e_2 e_1} & T_{e_3 e_2 e_2} & T_{e_3 e_2 e_3} \\
T_{e_3 e_3 e_1} & T_{e_3 e_3 e_2} & T_{e_3 e_3 e_3}
\end{pmatrix} \,.
\end{aligned}
\label{eqn:general_rank-3_tensor_in_e1e2e3_coordinates}
\end{equation}
Under an orthogonal transformation described by a matrix $R_{ij}$, \mbox{rank-3} tensors transform as
\begin{equation}
T_{ijk} \xrightarrow{R} |\mathbf{R}|^{N_\text{axial}} R_{i\ell}R_{jm}R_{kn}T_{\ell mn} \,,
\end{equation}
or, in matrix notation,
\begin{equation}
\mathbf{T}_i \xrightarrow{R} |\mathbf{R}|^{N_\text{axial}} R_{i\ell}\mathbf{R}\mathbf{T}_\ell\mathbf{R}^\mathrm{T} \,,
\end{equation}
where $\mathbf{T}_i$ is the matrix representation of $T_{ijk}$ for a given $i$.

$C_2^{e_2}$ symmetry imposes the constraints
\begin{equation}
\hspace{30.5pt}
\begin{aligned}
T_{e_1 e_1 e_1} = T_{e_3 e_3 e_3} &= 0 \,, \\
T_{e_1 e_1 e_3} = T_{e_1 e_3 e_1} = T_{e_3 e_1 e_1} &= 0 \,, \\
T_{e_3 e_3 e_1} = T_{e_3 e_1 e_3} = T_{e_1 e_3 e_3} &= 0 \,, \\
T_{e_1 e_2 e_2} = T_{e_2 e_1 e_2} = T_{e_2 e_2 e_1} &= 0 \,, \\
T_{e_3 e_2 e_2} = T_{e_2 e_3 e_2} = T_{e_2 e_2 e_3} &= 0 \,,
\end{aligned}
\qquad (C_2^{e_2})
\label{eqn:rank-3_zero-field_tensor_constraints-C2e2_symmetry}
\end{equation}
and $C_3^{e_3}$ symmetry imposes the constraints
\begin{equation}
\begin{aligned}
T_{e_1 e_1 e_1} = -T_{e_1 e_2 e_2} &= -T_{e_2 e_1 e_2} = -T_{e_2 e_2 e_1} \,, \\
T_{e_2 e_2 e_2} = -T_{e_2 e_1 e_1} &= -T_{e_1 e_2 e_1} = -T_{e_1 e_1 e_2} \,, \\
T_{e_1 e_1 e_3} &= T_{e_2 e_2 e_3} \,, \\
T_{e_1 e_3 e_1} &= T_{e_2 e_3 e_2} \,, \\
T_{e_3 e_1 e_1} &= T_{e_3 e_2 e_2} \,, \\
T_{e_1 e_2 e_3} &= -T_{e_2 e_1 e_3} \,, \\
T_{e_1 e_3 e_2} &= -T_{e_2 e_3 e_1} \,, \\
T_{e_3 e_2 e_1} &= -T_{e_3 e_1 e_2} \,, \\
T_{e_1 e_3 e_3} &= T_{e_3 e_1 e_3} = T_{e_3 e_3 e_1} = 0 \,, \\
T_{e_2 e_3 e_3} &= T_{e_3 e_2 e_3} = T_{e_3 e_3 e_2} = 0 \,.
\end{aligned}
(C_3^{e_3})
\label{eqn:rank-3_zero-field_tensor_constraints-C3e3_symmetry}
\end{equation}
Remarkably, for systems with $C_3^{e_3}$ symmetry, the fully \textit{longitudinal} components along the zigzag and armchair in-plane directions $e_1$ and $e_2$ (namely $T_{e_1 e_1 e_1}$ and $T_{e_2 e_2 e_2}$) are equal in magnitude to some partly \textit{transverse} components along these in-plane directions, as we can see in the first two lines in the equations above, namely
\begin{align}
T_{e_1 e_1 e_1} = -T_{e_1 e_2 e_2} &= -T_{e_2 e_1 e_2} = -T_{e_2 e_2 e_1} \,, \label{eqn:rank-3_zero-field_tensor_fully_longitudinal_equals_partly_transverse-top_line} \\
T_{e_2 e_2 e_2} = -T_{e_2 e_1 e_1} &= -T_{e_1 e_2 e_1} = -T_{e_1 e_1 e_2} \,. \label{eqn:rank-3_zero-field_tensor_fully_longitudinal_equals_partly_transverse-bottom_line}
\end{align}
For example, although one might have expected that $T_{e_2 e_2 e_2}$ and $T_{e_2 e_1 e_1}$ describe different physical processes and therefore have different values, $C_3^{e_3}$ symmetry nevertheless requires them to be the same.
We note that in systems that also have $C_2^{e_2}$ symmetry, the tensor components in Eq.~\ref{eqn:rank-3_zero-field_tensor_fully_longitudinal_equals_partly_transverse-top_line} (but not those in Eq.~\ref{eqn:rank-3_zero-field_tensor_fully_longitudinal_equals_partly_transverse-bottom_line}) will be zero, so we expect that in systems with small distortions that weakly break $C_2^{e_2}$ symmetry, the components in Eq.~\ref{eqn:rank-3_zero-field_tensor_fully_longitudinal_equals_partly_transverse-top_line} will be relatively small.

Unlike with \mbox{rank-2} zero-field tensors, \mbox{rank-3} zero-field tensors describing systems with $C_3^{e_3}$ symmetry do not generally have continuous rotational symmetry with respect to the axis perpendicular to the plane.
In fact, for systems with $C_3^{e_3}$ or $C_2^{e_2}$ symmetry, the in-plane zigzag and armchair directions $e_1$ and $e_2$ generally behave differently for \mbox{rank-3} zero-field tensors, so \mbox{rank-3} tensors are more sensitive at probing differences directional differences within the plane.

Finally, inversion symmetry again does not constrain the form of \mbox{rank-3} tensors, but it imposes that either none or two of the tensor indices correspond to the polar vectors, otherwise the tensor will equal zero.

The most general forms of \mbox{rank-3} zero-field tensors for the eight point groups generated by these three symmetries are given in \mbox{Table~\ref{tab:general_forms_of_tensors}}.

\subsubsection{Example: Thermomagnetic Susceptibility Tensor \texorpdfstring{$\chi^\text{thermomag}_{ijk}$}{chi thermomag ijk}}
An example of a \mbox{rank-3} zero-field tensor is the thermomagnetic susceptibility tensor
\begin{equation}
\chi^\text{thermomag}_{ijk} = \frac{\partial \kappa_{ij}}{\partial H_k}\biggr|_{\mathbf{H}=\mathbf{0}} \,,
\vspace{-3pt} 
\label{eqn:thermomagnetic_susceptibility_tensor}
\end{equation}
\begin{table*}[ht!]
\begin{center}
\begin{tabular}{|c|c|c|c|}
\hline
Point Groups
&
\multicolumn{1}{c|}{\begin{tabular}{c}\vspace{-2pt}Rank-2 Tensors\\(no external field)\vspace{1pt}\end{tabular}}
&
\multicolumn{1}{c|}{\begin{tabular}{c}\vspace{-2pt}Rank-3 Tensors\\(no external field)\vspace{1pt}\end{tabular}}
&
\multicolumn{1}{c|}{\begin{tabular}{c}\vspace{-2pt}Thermal Conductivity Tensor\\(in external magnetic field $\mathbf{H}$)\vspace{1pt}\end{tabular}}
\\ \hline
$1$,~~$\bar{1}$ & $T_{ij} = \begin{pmatrix}
A & B & C \\
D & E & F \\
G & I & J
\\ \vspace{-13.6pt} \hphantom{-B} & \hphantom{B} & \hphantom{-B}
\end{pmatrix}$ & \begin{tabular}{l@{}}
\vspace{-10pt} \\
$T_{e_1 jk}=\begin{pmatrix}
A_1 & B_1 & C_1 \\
D_1 & E_1 & F_1 \\
G_1 & I_1 & J_1
\\ \vspace{-13.6pt} \hphantom{-D} & \hphantom{-B} & \hphantom{-D}
\end{pmatrix}$ \vspace{5pt} \\
$T_{e_2 jk}=\begin{pmatrix}
A_2 & B_2 & C_2 \\
D_2 & E_2 & F_2 \\
G_2 & I_2 & J_2
\\ \vspace{-13.6pt} \hphantom{-D} & \hphantom{-B} & \hphantom{-D}
\end{pmatrix}$ \vspace{5pt} \\
$T_{e_3 jk}=\begin{pmatrix}
A_3 & B_3 & C_3 \\
D_3 & E_3 & F_3 \\
G_3 & I_3 & J_3
\\ \vspace{-13.6pt} \hphantom{-D} & \hphantom{-B} & \hphantom{-D}
\end{pmatrix}$ \\ \vspace{-10pt} \\
\end{tabular} & \begin{tabular}{l@{}}
\vspace{-11pt} \\
$\kappa_{ij}(H\hat{\mathbf{e}}_1)
=
\overset{\overset{\textbf{\small Even in $H$}}{\vphantom{,}}}{\begin{pmatrix}
A_1 & B_1 & C_1 \\
B_1 & D_1 & E_1 \\
C_1 & E_1 & F_1
\\ \vspace{-13.6pt} \hphantom{B_3} & \hphantom{D_3} & \hphantom{B_3}
\end{pmatrix}}
+
\overset{\overset{\textbf{\small Odd in $H$}}{\vphantom{,}}}{\begin{pmatrix}
0 & G_1 & I_1 \\
-G_1 & 0 & J_1 \\
-I_1 & -J_1 & 0
\\ \vspace{-13.6pt} \hphantom{-G_3} & \hphantom{-J_3} & \hphantom{-J_3}
\end{pmatrix}}$ \vspace{5pt} \\
$\kappa_{ij}(H\hat{\mathbf{e}}_2)
=
\begin{pmatrix}
A_2 & B_2 & C_2 \\
B_2 & D_2 & E_2 \\
C_2 & E_2 & F_2
\\ \vspace{-13.6pt} \hphantom{B_3} & \hphantom{D_3} & \hphantom{B_3}
\end{pmatrix}
+
\begin{pmatrix}
0 & G_2 & I_2 \\
-G_2 & 0 & J_2 \\
-I_2 & -J_2 & 0
\\ \vspace{-13.6pt} \hphantom{-G_3} & \hphantom{-J_3} & \hphantom{-J_3}
\end{pmatrix}$ \vspace{5pt} \\
$\kappa_{ij}(H\hat{\mathbf{e}}_3)
=
\begin{pmatrix}
A_3 & B_3 & C_3 \\
B_3 & D_3 & E_3 \\
C_3 & E_3 & F_3
\\ \vspace{-13.6pt} \hphantom{B_3} & \hphantom{D_3} & \hphantom{B_3}
\end{pmatrix}
+
\begin{pmatrix}
0 & G_3 & I_3 \\
-G_3 & 0 & J_3 \\
-I_3 & -J_3 & 0
\\ \vspace{-13.6pt} \hphantom{-G_3} & \hphantom{-J_3} & \hphantom{-J_3}
\end{pmatrix}$ \\ \vspace{-8pt} \\
\end{tabular}
\\ \hline
$2$,~~$2/m$ & $T_{ij} = \begin{pmatrix}
A & 0 & B \\
0 & C & 0 \\
D & 0 & E
\\ \vspace{-13.6pt} \hphantom{-B} & \hphantom{B} & \hphantom{-B}
\end{pmatrix}$ & \begin{tabular}{l@{}}
\vspace{-10pt} \\
$T_{e_1 jk}=\begin{pmatrix}
0 & A & 0 \\
B & 0 & C \\
0 & D & 0
\\ \vspace{-13.6pt} \hphantom{-D} & \hphantom{-B} & \hphantom{-D}
\end{pmatrix}$ \vspace{5pt} \\
$T_{e_2 jk}=\begin{pmatrix}
E & 0 & F \\
0 & G & 0 \\
I & 0 & J
\\ \vspace{-13.6pt} \hphantom{-D} & \hphantom{-B} & \hphantom{-D}
\end{pmatrix}$ \vspace{5pt} \\
$T_{e_3 jk}=\begin{pmatrix}
0 & K & 0 \\
L & 0 & M \\
0 & N & 0
\\ \vspace{-13.6pt} \hphantom{-D} & \hphantom{-B} & \hphantom{-D}
\end{pmatrix}$ \\ \vspace{-10pt} \\
\end{tabular} & \begin{tabular}{l@{}}
\vspace{-11pt} \\
$\kappa_{ij}(H\hat{\mathbf{e}}_1)
=
\overset{\overset{\textbf{\small Even in $H$}}{\vphantom{,}}}{\begin{pmatrix}
A & 0 & B \\
0 & C & 0 \\
B & 0 & D
\\ \vspace{-13.6pt} \hphantom{B_3} & \hphantom{D_3} & \hphantom{B_3}
\end{pmatrix}}
+
\overset{\overset{\textbf{\small Odd in $H$}}{\vphantom{,}}}{\begin{pmatrix}
0 & E & 0 \\
-E & 0 & F \\
0 & -F & 0
\\ \vspace{-13.6pt} \hphantom{-G_3} & \hphantom{-J_3} & \hphantom{-J_3}
\end{pmatrix}}$ \vspace{5pt} \\
$\kappa_{ij}(H\hat{\mathbf{e}}_2)
=
\begin{pmatrix}
G & 0 & I \\
0 & J & 0 \\
I & 0 & K
\\ \vspace{-13.6pt} \hphantom{B_3} & \hphantom{D_3} & \hphantom{B_3}
\end{pmatrix}
+
\begin{pmatrix}
0 & 0 & L \\
0 & 0 & 0 \\
-L & 0 & 0
\\ \vspace{-13.6pt} \hphantom{-G_3} & \hphantom{-J_3} & \hphantom{-J_3}
\end{pmatrix}$ \vspace{5pt} \\
$\kappa_{ij}(H\hat{\mathbf{e}}_3)
=
\begin{pmatrix}
M & 0 & N \\
0 & P & 0 \\
N & 0 & Q
\\ \vspace{-13.6pt} \hphantom{B_3} & \hphantom{D_3} & \hphantom{B_3}
\end{pmatrix}
+
\begin{pmatrix}
0 & R & 0 \\
-R & 0 & S \\
0 & -S & 0
\\ \vspace{-13.6pt} \hphantom{-G_3} & \hphantom{-J_3} & \hphantom{-J_3}
\end{pmatrix}$ \\ \vspace{-8pt} \\
\end{tabular}
\\ \hline
$3$,~~$\bar{3}$ & $T_{ij} = \begin{pmatrix}
A & B & 0 \\
-B & A & 0 \\
0 & 0 & C
\\ \vspace{-13.6pt} \hphantom{-B} & \hphantom{B} & \hphantom{-B}
\end{pmatrix}$ & \begin{tabular}{l@{}}
\vspace{-10pt} \\
$T_{e_1 jk}=\begin{pmatrix}
A & B & C \\
B & -A & D \\
E & F & 0
\\ \vspace{-13.6pt} \hphantom{-D} & \hphantom{-B} & \hphantom{-D}
\end{pmatrix}$ \vspace{5pt} \\
$T_{e_2 jk}=\begin{pmatrix}
B & -A & -D \\
-A & -B & C \\
-F & E & 0
\\ \vspace{-13.6pt} \hphantom{-D} & \hphantom{-B} & \hphantom{-D}
\end{pmatrix}$ \vspace{5pt} \\
$T_{e_3 jk}=\begin{pmatrix}
G & I & 0 \\
-I & G & 0 \\
0 & 0 & J
\\ \vspace{-13.6pt} \hphantom{-D} & \hphantom{-B} & \hphantom{-D}
\end{pmatrix}$ \\ \vspace{-10pt} \\
\end{tabular} & \begin{tabular}{l@{}}
\vspace{-11pt} \\
$\kappa_{ij}(H\hat{\mathbf{e}}_1)
=
\overset{\overset{\textbf{\small Even in $H$}}{\vphantom{,}}}{\begin{pmatrix}
A & B & C \\
B & D & E \\
C & E & F
\\ \vspace{-13.6pt} \hphantom{B_3} & \hphantom{D_3} & \hphantom{B_3}
\end{pmatrix}}
+
\overset{\overset{\textbf{\small Odd in $H$}}{\vphantom{,}}}{\begin{pmatrix}
0 & G & I \\
-G & 0 & J \\
-I & -J & 0
\\ \vspace{-13.6pt} \hphantom{-G_3} & \hphantom{-J_3} & \hphantom{-J_3}
\end{pmatrix}}$ \vspace{5pt} \\
$\kappa_{ij}(H\hat{\mathbf{e}}_2)
=
\begin{pmatrix}
K & L & M \\
L & N & P \\
M & P & Q
\\ \vspace{-13.6pt} \hphantom{B_3} & \hphantom{D_3} & \hphantom{B_3}
\end{pmatrix}
+
\begin{pmatrix}
0 & R & S \\
-R & 0 & T \\
-S & -T & 0
\\ \vspace{-13.6pt} \hphantom{-G_3} & \hphantom{-J_3} & \hphantom{-J_3}
\end{pmatrix}$ \vspace{5pt} \\
$\kappa_{ij}(H\hat{\mathbf{e}}_3)
=
\begin{pmatrix}
U & 0 & 0 \\
0 & U & 0 \\
0 & 0 & V
\\ \vspace{-13.6pt} \hphantom{B_3} & \hphantom{D_3} & \hphantom{B_3}
\end{pmatrix}
+
\begin{pmatrix}
0 & W & 0 \\
-W & 0 & 0 \\
0 & 0 & 0
\\ \vspace{-13.6pt} \hphantom{-G_3} & \hphantom{-J_3} & \hphantom{-J_3}
\end{pmatrix}$ \\ \vspace{-8pt} \\
\end{tabular}
\\ \hline
$32$,~~$\bar{3}m$ & $T_{ij} = \begin{pmatrix}
A & 0 & 0 \\
0 & A & 0 \\
0 & 0 & B
\\ \vspace{-13.6pt} \hphantom{-B} & \hphantom{B} & \hphantom{-B}
\end{pmatrix}$ & \begin{tabular}{l@{}}
\vspace{-10pt} \\
$T_{e_1 jk}=\begin{pmatrix}
0 & A & 0 \\
A & 0 & B \\
0 & C & 0
\\ \vspace{-13.6pt} \hphantom{-D} & \hphantom{-B} & \hphantom{-D}
\end{pmatrix}$ \vspace{5pt} \\
$T_{e_2 jk}=\begin{pmatrix}
A & 0 & -B \\
0 & -A & 0 \\
-C & 0 & 0
\\ \vspace{-13.6pt} \hphantom{-D} & \hphantom{-B} & \hphantom{-D}
\end{pmatrix}$ \vspace{5pt} \\
$T_{e_3 jk}=\begin{pmatrix}
0 & D & 0 \\
-D & 0 & 0 \\
0 & 0 & 0
\\ \vspace{-13.6pt} \hphantom{-D} & \hphantom{-B} & \hphantom{-D}
\end{pmatrix}$ \\ \vspace{-10pt} \\
\end{tabular} & \begin{tabular}{l@{}}
\vspace{-11pt} \\
$\kappa_{ij}(H\hat{\mathbf{e}}_1)
=
\overset{\overset{\textbf{\small Even in $H$}}{\vphantom{,}}}{\begin{pmatrix}
A & 0 & B \\
0 & C & 0 \\
B & 0 & D
\\ \vspace{-13.6pt} \hphantom{B_3} & \hphantom{D_3} & \hphantom{B_3}
\end{pmatrix}}
+
\overset{\overset{\textbf{\small Odd in $H$}}{\vphantom{,}}}{\begin{pmatrix}
0 & E & 0 \\
-E & 0 & F \\
0 & -F & 0
\\ \vspace{-13.6pt} \hphantom{-G_3} & \hphantom{-J_3} & \hphantom{-J_3}
\end{pmatrix}}$ \vspace{5pt} \\
$\kappa_{ij}(H\hat{\mathbf{e}}_2)
=
\begin{pmatrix}
G & 0 & I \\
0 & J & 0 \\
I & 0 & K
\\ \vspace{-13.6pt} \hphantom{B_3} & \hphantom{D_3} & \hphantom{B_3}
\end{pmatrix}
+
\begin{pmatrix}
0 & 0 & L \\
0 & 0 & 0 \\
-L & 0 & 0
\\ \vspace{-13.6pt} \hphantom{-G_3} & \hphantom{-J_3} & \hphantom{-J_3}
\end{pmatrix}$ \vspace{5pt} \\
$\kappa_{ij}(H\hat{\mathbf{e}}_3)
=
\begin{pmatrix}
M & 0 & 0 \\
0 & M & 0 \\
0 & 0 & N
\\ \vspace{-13.6pt} \hphantom{B_3} & \hphantom{D_3} & \hphantom{B_3}
\end{pmatrix}
+
\begin{pmatrix}
0 & P & 0 \\
-P & 0 & 0 \\
0 & 0 & 0
\\ \vspace{-13.6pt} \hphantom{-G_3} & \hphantom{-J_3} & \hphantom{-J_3}
\end{pmatrix}$ \\ \vspace{-8pt} \\
\end{tabular}
\\ \hline
\end{tabular}
\caption{General forms of \mbox{rank-2} and \mbox{rank-3} tensors in systems with no external field (magnetic or electric) and of the thermal conductivity tensor $\kappa_{ij}(\mathbf{H})$ in systems with an external magnetic field $\mathbf{H}$ along the high-symmetry directions \mbox{$\alpha=e_1,e_2,e_3$} \mbox{($e_1$ = in-plane zigzag direction, $e_2$ = in-plane armchair direction, $e_3$ = out-of-plane direction; see Fig.~\ref{fig:honeycomb_lattice_of_edge-sharing_octahedra})} for systems of various point groups.
The components of these tensors are expressed in $e_1 e_2 e_3$ coordinates (e.g., see Eqs.~\ref{eqn:general_rank-2_tensor_in_e1e2e3_coordinates} and \ref{eqn:general_rank-3_tensor_in_e1e2e3_coordinates}).
The components of the thermal conductivity tensor that are even functions of $H$ correspond to the thermomagnetic conductivity, whereas those that are odd functions of $H$ correspond to the thermal Hall conductivity.}
\label{tab:general_forms_of_tensors}
\end{center}
\end{table*}
\clearpage
\noindent where $\kappa_{ij}$ is the thermal conductivity tensor, defined by
\begin{equation}
(J_Q)_i = -\kappa_{ij} \nabla_j T \,,
\end{equation}
$\mathbf{J}_Q$ is the heat current, $\boldsymbol{\nabla} T$ is the temperature gradient, and $\mathbf{H}$ is the external magnetic field.
Even though we are taking a magnetic field derivative, this is still a zero-field tensor because we are evaluating the derivative in the zero-field limit (i.e., the infinitesimally small field is only being used to probe the zero-field ground state).
Also note that while $\chi^\text{thermomag}_{ijk}$ is linear in the vectors $\mathbf{H}$ and $\boldsymbol{\nabla} T$,
\mbox{rank-3} tensors can also be quadratic in a given vector, such as the nonlinear magnetic susceptibility tensor~\cite{Shivaram2014_RevSciInstrum,Shivaram2014_PRB,Shivaram2017,Shivaram2018}
\begin{equation}
\chi^\text{nonlinear}_{ijk} = \frac{\partial^2 M_i}{\partial H_j \partial H_k} \,,
\end{equation}
which is quadratic in $\mathbf{H}$.

\begin{figure}[ht!]
\begin{center}
\includegraphics[width=2.6in]{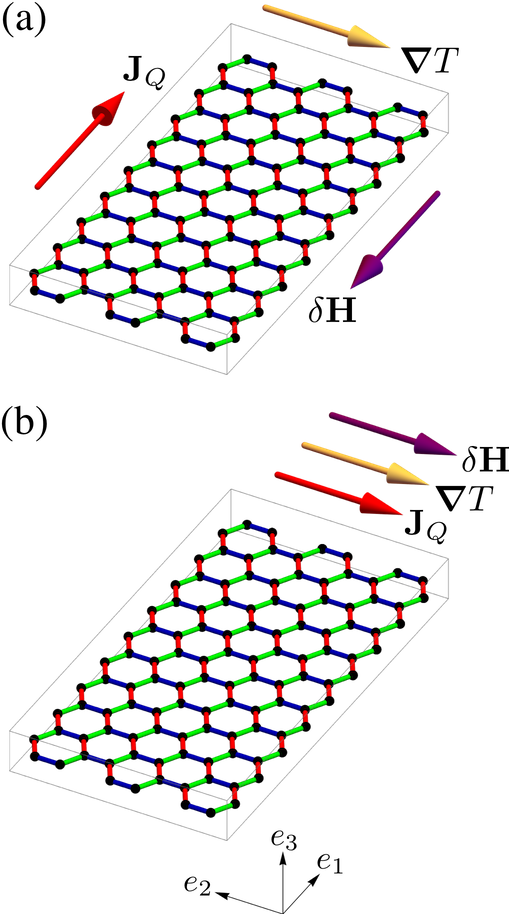}
\caption{Example illustrating one of the unusual equalities for \mbox{rank-3} tensors describing crystals with $C_3^{e_3}$ symmetry (i.e., belonging to the trigonal point groups $3$, $\bar{3}$, $32$, or $\bar{3}m$).
In the absence of external magnetic or electric fields,
(a) the change in the thermal Hall conductivity $\delta\kappa_{e_1 e_2}$ that results from applying a small magnetic field $\delta\mathbf{H}$ along the zigzag direction $e_1$
will be the same as
(b) the change in the longitudinal thermal conductivity $\delta\kappa_{e_2 e_2}$ that results from applying a small magnetic field $\delta\mathbf{H}$ along the armchair direction $e_2$
(Eq.~\ref{eqn:thermomagnetic_susceptibility-surprising_equality}).
$\mathbf{J}_Q$ is the heat current and $\boldsymbol{\nabla} T$ is the temperature gradient.
Reversing the direction of any one of the three vectors simply reverses the sign of the change in the thermal conductivity.
These results hold when the system is either not magnetized, or magnetized along the out-of-plane direction ($e_3$).
Applying Eqs.~\ref{eqn:rank-3_zero-field_tensor_fully_longitudinal_equals_partly_transverse-top_line} and \ref{eqn:rank-3_zero-field_tensor_fully_longitudinal_equals_partly_transverse-bottom_line} using the thermomagnetic susceptibility tensor
\mbox{$\chi^\text{thermomag}_{ijk}=(\partial \kappa_{ij}/\partial H_k)|_{\mathbf{H}=\mathbf{0}}$}
yields several other similar equalities,
illustrated in Fig.~\ref{fig:thermomagnetic_susceptibility-all_predictions}.}
\label{fig:thermomagnetic_susceptibility-one_example_of_fully_longitudinal_equals_partly_transverse}
\end{center}
\end{figure}

For a material with $C_3^{e_3}$ symmetry, such as CrI$_3$ in the rhombohedral configuration $R\bar{3}$~\cite{McGuire2015,Ubrig2020}, we expect that
\mbox{$\chi^\text{thermomag}_{e_1 e_2 e_1}=-\chi^\text{thermomag}_{e_2 e_2 e_2}$} (see Eq.~\ref{eqn:rank-3_zero-field_tensor_fully_longitudinal_equals_partly_transverse-bottom_line}), or more explicitly,
\begin{equation}
\frac{\partial \kappa_{e_1 e_2}}{\partial H_{e_1}}\biggr|_{\mathbf{H}=\mathbf{0}}
=
-\frac{\partial \kappa_{e_2 e_2}}{\partial H_{e_2}}\biggr|_{\mathbf{H}=\mathbf{0}} \,.
\label{eqn:thermomagnetic_susceptibility-surprising_equality}
\end{equation}
This result is surprising because the left side corresponds to the field derivative of a thermal Hall conductivity \mbox{($\mathbf{J}_Q \perp \boldsymbol{\nabla} T$)}~(Fig.~\ref{fig:thermomagnetic_susceptibility-one_example_of_fully_longitudinal_equals_partly_transverse}a),
whereas the right side corresponds to the field derivative of a longitudinal thermal conductivity \mbox{($\mathbf{J}_Q \parallel \boldsymbol{\nabla} T$)}~(Fig.~\ref{fig:thermomagnetic_susceptibility-one_example_of_fully_longitudinal_equals_partly_transverse}b).
These results still hold when the system is magnetized along the out-of-plane direction, as this does not break $C_3^{e_3}$ symmetry.
We can obtain several other similar expressions using Eqs.~\ref{eqn:rank-3_zero-field_tensor_fully_longitudinal_equals_partly_transverse-top_line} and \ref{eqn:rank-3_zero-field_tensor_fully_longitudinal_equals_partly_transverse-bottom_line}.

Note that even though components such as $\chi^\text{thermomag}_{e_3 e_1 e_2}$ and $\chi^\text{thermomag}_{e_1 e_3 e_2}$ are always allowed to be nonzero for all eight point groups possible, we nevertheless expect them to be zero for monolayer systems, since heat currents and temperature gradients cannot physically be oriented perpendicular to a 2D system~\footnote{Of course, the component $\chi^\text{thermomag}_{e_1 e_2 e_3}$ can still be nonzero in monolayer systems, since the magnetic field can be oriented perpendicular to the honeycomb plane, as is commonly the case thermal Hall experiments~\cite{KasaharaOhnishi2018}}.
We therefore only expect these components to become relevant for bulk systems.

\subsection{Rank-\texorpdfstring{$r$}{r} Tensors}
Under a transformation described by an orthogonal transformation matrix $R_{ij}$, a general \mbox{rank-$r$} tensor $T_{i_1 i_2 \hdots i_r}$ transforms as
\begin{align}
T_{i_1 i_2 \hdots i_r} &\xrightarrow{R} |\mathbf{R}|^{N_\text{axial}} \left(\prod_{n=1}^r R_{i_n j_n}\right) T_{j_1 j_2 \hdots j_r} \nonumber \\
&\hspace{4.655pt}= |\mathbf{R}|^{N_\text{axial}} R_{i_1 j_1} R_{i_2 j_2} \cdots R_{i_r j_r} T_{j_1 j_2 \hdots j_r} \,.
\end{align}

Under a $C_2^{e_2}$ transformation, the components of a \mbox{rank-$r$} tensor transform as
\begin{equation}
T_{i_1 i_2 \hdots i_r} \xrightarrow{C_2^{e_2}} (-1)^{N_{e_1} + N_{e_3}} T_{i_1 i_2 \hdots i_r} \,,
\end{equation}
where $N_\alpha$
\mbox{($\alpha=e_1,e_2,e_3$)}
is the number of indices in \mbox{$\{i_1, i_2, \hdots, i_r\}$} equal to $\alpha$~\footnote{For example, for the \mbox{rank-3} tensor component $T_{e_1 e_1 e_3}$, \mbox{$N_{e_1}=2$} and \mbox{$N_{e_3}=1$}.}.
Invariance under $C_2^{e_2}$ therefore implies that
\begin{equation}
T_{i_1 i_2 \hdots i_r} = 0 \quad \text{if} \quad N_{e_1} + N_{e_3} \quad \text{is odd} \,. \quad (C_2^{e_2})
\end{equation}

Under a $C_3^{e_3}$ transformation, the components of a \mbox{rank-$r$} tensor do not transform in a straightforward manner due to the mixing of the $e_1$ and $e_2$ directions.
We therefore do not have a simple generalization of the constraints this symmetry places on tensors of any rank.

Finally, under inversion, \mbox{rank-$r$} tensors transform as
\begin{equation}
T_{i_1 i_2 \hdots i_r} \xrightarrow{\mathcal{I}} (-1)^{N_\text{polar}} T_{i_1 i_2 \hdots i_r} \,,
\end{equation}
where $N_\text{polar}$
is the number of indices in $T_{i_1 i_2 \hdots i_r}$ corresponding to polar vectors~\footnote{Throughout this paper, we will assume that each tensor index transforms as either a polar vector or an axial vector.
We will therefore not consider tensors of the form \mbox{$T_{ij}=\partial A_i/\partial B_j + \partial B_i/\partial A_j$} for
$\mathbf{A}$ polar and $\mathbf{B}$ axial,
for example, as these are just linear combinations of the types of tensors we will consider.}.
Invariance under inversion therefore implies that
\begin{equation}
T_{i_1 i_2 \hdots i_r} = 0 \quad \text{if} \quad N_\text{polar} \quad \text{is odd} \,. \quad (\mathcal{I})
\end{equation}

In \mbox{Table~\ref{tab:constraints_imposed_by_crystal_symmetries}} we summarize the constraints placed on zero-field tensors by the three possible generating symmetries of the honeycomb lattice of edge-sharing octahedra.

\section{Field-Dependent Tensors}
\label{section:field-dependent_tensors}
In this section we describe the types of symmetry constraints placed on tensors for systems in an external magnetic or electric field, and as an example we obtain the symmetry constraints on the magnetic-field-dependent thermal conductivity tensor $\kappa_{ij}(\mathbf{H})$.

Tensors that depend on a magnetic or electric field $\mathbf{F}$ are constrained using the Grabner--Swanson symmetry constraint equation~\cite{GrabnerSwanson1962,AkgozSaunders1975_pt_I,AkgozSaunders1975_pt_II}
\begin{equation}
T_{i_1 i_2 \hdots i_r}(\mathbf{F}) = T'_{i_1 i_2 \hdots i_r}(\tilde{\mathbf{F}})
\label{eqn:Grabner-Swanson_symmetry_constraint}
\end{equation}
for each coordinate transformation matrix $\boldsymbol{\mathcal{S}}$ corresponding to a crystallographic symmetry,
where \mbox{$T'_{i_1 i_2 \hdots i_r}(\tilde{\mathbf{F}})=|\boldsymbol{\mathcal{S}}|^{N_\text{axial}} \mathcal{S}_{i_1 j_1} \mathcal{S}_{i_2 j_2} \cdots \mathcal{S}_{i_r j_r} T_{j_1 j_2 \hdots j_r}(\tilde{\mathbf{F}})$} is the original tensor expressed in the transformed coordinates (passive transformation);
\mbox{$\tilde{\mathbf{F}}=|\boldsymbol{\mathcal{S}}|^{\delta_\text{axial}}\boldsymbol{\mathcal{S}}^\mathrm{T}\mathbf{F}$} is the transformed field expressed in the original coordinates (active transformation);
and $\delta_\text{axial}$
is 1 if $\mathbf{F}$ is an axial vector, and 0 if it is a polar vector~\footnote{For systems with more than one external field \mbox{($\mathbf{F}_1$, $\mathbf{F}_2$, $\hdots$)}, the Grabner--Swanson equation~(Eq.~\ref{eqn:Grabner-Swanson_symmetry_constraint}) generalizes to \mbox{$T_{i_1 i_2 \hdots i_r}(\mathbf{F}_1,\mathbf{F}_2,\hdots) = T'_{i_1 i_2 \hdots i_r}(\tilde{\mathbf{F}}_1,\tilde{\mathbf{F}}_2,\hdots$}).}.
Note that we are not using the notation $\tilde{\mathbf{F}}$ for the actively transformed field because $\mathbf{F}'$ corresponds to the original, untransformed field expressed in the transformed coordinates, whereas $\tilde{\mathbf{F}}$ corresponds to the transformed field expressed in the original coordinates.

\begin{figure}[ht!]
\begin{center}
\includegraphics[width=3in]{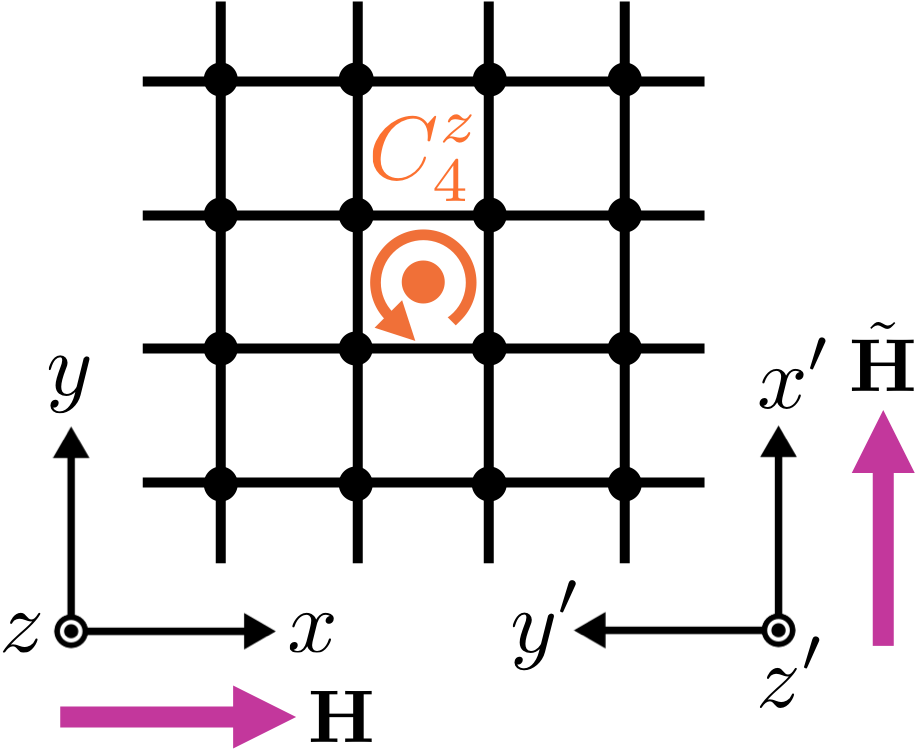}
\caption{Example illustrating the type of symmetry constraint imposed by the Grabner--Swanson equation~(Eq.~\ref{eqn:Grabner-Swanson_symmetry_constraint}) on systems in an external magnetic or electric field.
For a square lattice with four-fold ($90^\circ$) rotational symmetry with respect to the $z$-axis ($C_4^z$) in the presence of an external magnetic field $\mathbf{H}$ applied along the $x$-axis,
even though the magnetic field breaks the system's $C_4^z$ symmetry, we can nevertheless use this symmetry to state that for a \mbox{rank-2} field-dependent tensor $T_{ij}(\mathbf{H})$, we must have \mbox{$T_{ij}(\mathbf{H}) = T'_{ij}(\tilde{\mathbf{H}})$}, where $T'_{ij}(\tilde{\mathbf{H}})$ is the tensor expressed in the transformed coordinates (passive transformation) and
$\tilde{\mathbf{H}}$ is the transformed field (active transformation). For the element $T_{xz}(H\hat{\mathbf{x}})$, for example, this gives \mbox{$T_{xz}(H\hat{\mathbf{x}}) = T_{yz}(H\hat{\mathbf{y}})$}.}
\label{fig:Grabner-Swanson_symmetry_constraint_example}
\end{center}
\end{figure}

For example, consider a magnetic-field-dependent \mbox{rank-2} response tensor $T_{ij}(\mathbf{H})$ describing a square lattice with four-fold ($90^\circ$) rotational symmetry along the $z$-axis ($C_4^z$) in the presence of an external magnetic field in the $x$ direction \mbox{($\mathbf{H}=H\hat{\mathbf{x}}$)}, as shown in Fig.~\ref{fig:Grabner-Swanson_symmetry_constraint_example}.
Under $C_4^z$, the coordinate system $xyz$ will rotate counterclockwise by $90^\circ$ with respect to the $z$-axis, giving the transformed coordinates $x'y'z'$.
The passively transformed tensor $T'_{ij}$ is expressed in terms of these transformed coordinates.
The actively transformed field $\tilde{\mathbf{H}}$ is similarly obtained by rotating $\mathbf{H}$ counterclockwise by $90^\circ$ with respect to the $z$-axis, giving \mbox{$\tilde{\mathbf{H}}=H\hat{\mathbf{x}}'=H\hat{\mathbf{y}}$}.
Applying the Grabner--Swanson equation on the $xz$ element of this rank-2 tensor gives
\begin{align}
T_{xz}(H\hat{\mathbf{x}}) &= T_{x'z'}(H\hat{\mathbf{x}}') \\
&= T_{yz}(H\hat{\mathbf{y}}) \,,
\end{align}
where \mbox{$T_{xz}(H\hat{\mathbf{x}})$} describes the $xz$ response when $\mathbf{H}$ points along $\hat{\mathbf{x}}$ and
\mbox{$T_{x'z'}(H\hat{\mathbf{x}}')$} describes the $x'z'$ response when $\mathbf{H}$ points along $\hat{\mathbf{x}}'$.
Since the $x'$ direction is the same as the $y$ direction, and the $z'$ direction is the same as the $z$ direction~(Fig.~\ref{fig:Grabner-Swanson_symmetry_constraint_example}),
\mbox{$T_{x'z'}(H\hat{\mathbf{x}}')$} is therefore equal to
\mbox{$T_{yz}(H\hat{\mathbf{y}})$}, which describes the $yz$ response when $\mathbf{H}$ points along $\hat{\mathbf{y}}$.

\vspace{-20pt} 
\subsection{Example: Thermal Conductivity Tensor \texorpdfstring{$\kappa_{ij}(\mathbf{H})$}{kappa ij(H)}}
An example of a field-dependent \mbox{rank-2} tensor is the thermal conductivity tensor $\kappa_{ij}(\mathbf{H})$, given by
\begin{equation}
(J_Q)_i = -\kappa_{ij}(\mathbf{H}) \nabla_j T \,,
\label{eqn:thermal_conductivity_tensor}
\end{equation}
where $\mathbf{J}_Q$ is the heat current, $\boldsymbol{\nabla} T$ is the temperature gradient, and $\mathbf{H}$ is the external magnetic field.
It is useful to express it as a sum of even and odd functions of the magnetic field,
\begin{equation}
\kappa_{ij}(\mathbf{H}) = \kappa_{ij}^\text{even}(\mathbf{H}) + \kappa_{ij}^\text{odd}(\mathbf{H}) \,,
\label{eqn:kappa_ij_even+odd_index_form}
\end{equation}
where \mbox{$\kappa_{ij}^\text{even}(\mathbf{H})=\kappa_{ij}^\text{even}(-\mathbf{H})$} and $\kappa_{ij}^\text{odd}(\mathbf{H})=-\kappa_{ij}^\text{odd}(-\mathbf{H})$,
and are experimentally obtained by reversing the direction of the applied field~\cite{AkgozSaunders1975_pt_I,AkgozSaunders1975_pt_II}:
\begin{align}
\kappa_{ij}^\text{even}(\mathbf{H}) &= \frac{1}{2}[\kappa_{ij}(\mathbf{H}) + \kappa_{ij}(-\mathbf{H})] \,, \\ \kappa_{ij}^\text{odd}(\mathbf{H}) &= \frac{1}{2}[\kappa_{ij}(\mathbf{H}) - \kappa_{ij}(-\mathbf{H})] \,.
\end{align}
Since $\kappa_{ij}(\mathbf{H})$ satisfies the Onsager relation~\cite{Onsager1931,AkgozSaunders1975_pt_I,AkgozSaunders1975_pt_II}
\begin{equation}
\kappa_{ij}(\mathbf{H}) = \kappa_{ji}(-\mathbf{H}) \,,
\end{equation}
then $\kappa_{ij}^\text{even}(\mathbf{H})$ must be a symmetric tensor and $\kappa_{ij}^\text{odd}(\mathbf{H})$ must be antisymmetric.
In matrix form, Eq.~\ref{eqn:kappa_ij_even+odd_index_form} is therefore an equation of the general form
\begin{equation}
\kappa_{ij}(\mathbf{H}) = 
\overset{\overset{\textbf{\small Even in $\mathbf{H}$}}{\vphantom{,}}}{
\underbrace{
\begin{pmatrix}
A & B & C \\
B & D & E \\
C & E & F
\end{pmatrix}
}_{\kappa_{ij}^\text{even}(\mathbf{H})}
}
+
\overset{\overset{\textbf{\small Odd in $\mathbf{H}$}}{\vphantom{,}}}{
\underbrace{
\begin{pmatrix}
0 & G & I \\
-G & 0 & J \\
-I & -J & 0
\end{pmatrix}
}_{\kappa_{ij}^\text{odd}(\mathbf{H})}
}
\label{eqn:kappa_ij_even+odd_matrix_form}
\end{equation}
before placing any crystal symmetry constraints.
Following Ref.~\cite{AkgozSaunders1975_pt_II}, we identify $\kappa_{ij}^\text{even}(\mathbf{H})$ as the \mbox{\textit{thermomagnetic conductivity}} and $\kappa_{ij}^\text{odd}(\mathbf{H})$ as the \mbox{\textit{thermal Hall conductivity}}.

We will now obtain the symmetry constraints for $\kappa_{ij}(\mathbf{H})$ for the cases where the magnetic field points along a zigzag direction ($e_1$), an armchair direction ($e_2$), or the direction perpendicular to the plane ($e_3$).
Applying the Grabner--Swanson equation~(Eq.~\ref{eqn:Grabner-Swanson_symmetry_constraint}) using the symmetry $C_2^{e_2}$ for the case where the magnetic field points along the zigzag direction $e_1$ gives
\begin{align}
&
\begin{pmatrix}
\kappa_{e_1 e_1}(H\hat{\mathbf{e}}_1) & \kappa_{e_1 e_2}(H\hat{\mathbf{e}}_1) & \kappa_{e_1 e_3}(H\hat{\mathbf{e}}_1) \\
\kappa_{e_2 e_1}(H\hat{\mathbf{e}}_1) & \kappa_{e_2 e_2}(H\hat{\mathbf{e}}_1) & \kappa_{e_2 e_3}(H\hat{\mathbf{e}}_1) \\
\kappa_{e_3 e_1}(H\hat{\mathbf{e}}_1) & \kappa_{e_3 e_2}(H\hat{\mathbf{e}}_1) & \kappa_{e_3 e_3}(H\hat{\mathbf{e}}_1)
\\ \vspace{-13.6pt} \hphantom{-\kappa_{e_2 e_1}(-H\hat{\mathbf{e}}_1)} & \hphantom{-\kappa_{e_3 e_2}(-H\hat{\mathbf{e}}_1)} & \hphantom{-\kappa_{e_2 e_3}(-H\hat{\mathbf{e}}_1)}
\end{pmatrix} \nonumber \\
&= \begin{pmatrix}
\kappa_{e_1 e_1}(-H\hat{\mathbf{e}}_1) & -\kappa_{e_1 e_2}(-H\hat{\mathbf{e}}_1) & \kappa_{e_1 e_3}(-H\hat{\mathbf{e}}_1) \\
-\kappa_{e_2 e_1}(-H\hat{\mathbf{e}}_1) & \kappa_{e_2 e_2}(-H\hat{\mathbf{e}}_1) & -\kappa_{e_2 e_3}(-H\hat{\mathbf{e}}_1) \\
\kappa_{e_3 e_1}(-H\hat{\mathbf{e}}_1) & -\kappa_{e_3 e_2}(-H\hat{\mathbf{e}}_1) & \kappa_{e_3 e_3}(-H\hat{\mathbf{e}}_1)
\end{pmatrix} \,,
\end{align}
which constrains $\kappa_{ij}(H\hat{\mathbf{e}}_1)$ to be of the form
\begin{equation}
\kappa_{ij}(H\hat{\mathbf{e}}_1) =
\overset{\overset{\textbf{\small Even in $H$}}{\vphantom{,}}}{\begin{pmatrix}
A & 0 & B \\
0 & C & 0 \\
B & 0 & D
\end{pmatrix}}
+
\overset{\overset{\textbf{\small Odd in $H$}}{\vphantom{,}}}{\begin{pmatrix}
0 & E & 0 \\
-E & 0 & F \\
0 & -F & 0
\end{pmatrix}} \,, \quad (C_2^{e_2}) \label{eqn:kappa_ij(He1)-constrained_by_C2e2}
\end{equation}
and similarly for when the field points perpendicular to the plane (i.e., along $e_3$).
However, if the magnetic field points along the armchair direction $e_2$, the Grabner--Swanson equation gives
\begin{align}
&
\begin{pmatrix}
\kappa_{e_1 e_1}(H\hat{\mathbf{e}}_2) & \kappa_{e_1 e_2}(H\hat{\mathbf{e}}_2) & \kappa_{e_1 e_3}(H\hat{\mathbf{e}}_2) \\
\kappa_{e_2 e_1}(H\hat{\mathbf{e}}_2) & \kappa_{e_2 e_2}(H\hat{\mathbf{e}}_2) & \kappa_{e_2 e_3}(H\hat{\mathbf{e}}_2) \\
\kappa_{e_3 e_1}(H\hat{\mathbf{e}}_2) & \kappa_{e_3 e_2}(H\hat{\mathbf{e}}_2) & \kappa_{e_3 e_3}(H\hat{\mathbf{e}}_2)
\\ \vspace{-13.6pt} \hphantom{-\kappa_{e_2 e_1}(H\hat{\mathbf{e}}_2)} & \hphantom{-\kappa_{e_3 e_2}(H\hat{\mathbf{e}}_2)} & \hphantom{-\kappa_{e_2 e_3}(H\hat{\mathbf{e}}_2)}
\end{pmatrix} \nonumber \\
&= \begin{pmatrix}
\kappa_{e_1 e_1}(H\hat{\mathbf{e}}_2) & -\kappa_{e_1 e_2}(H\hat{\mathbf{e}}_2) & \kappa_{e_1 e_3}(H\hat{\mathbf{e}}_2) \\
-\kappa_{e_2 e_1}(H\hat{\mathbf{e}}_2) & \kappa_{e_2 e_2}(H\hat{\mathbf{e}}_2) & -\kappa_{e_2 e_3}(H\hat{\mathbf{e}}_2) \\
\kappa_{e_3 e_1}(H\hat{\mathbf{e}}_2) & -\kappa_{e_3 e_2}(H\hat{\mathbf{e}}_2) & \kappa_{e_3 e_3}(H\hat{\mathbf{e}}_2)
\end{pmatrix} \,,
\end{align}
which constrains $\kappa_{ij}(H\hat{\mathbf{e}}_2)$ to be of the form
\begin{equation}
\kappa_{ij}(H\hat{\mathbf{e}}_2) =
\overset{\overset{\textbf{\small Even in $H$}}{\vphantom{,}}}{\begin{pmatrix}
G & 0 & I \\
0 & J & 0 \\
I & 0 & K
\end{pmatrix}}+
\overset{\overset{\textbf{\small Odd in $H$}}{\vphantom{,}}}{\begin{pmatrix}
0 & 0 & L \\
0 & 0 & 0 \\
-L & 0 & 0
\\ \vspace{-13.6pt} \hphantom{-Q} & \hphantom{-R} & \hphantom{R}
\end{pmatrix}} \,. \quad (C_2^{e_2}) \label{eqn:kappa_ij(He2)-constrained_by_C2e2}
\end{equation}
The symmetry $C_3^{e_3}$ does not constrain the form of $\kappa_{ij}(\mathbf{H})$ when the magnetic field points along the zigzag or armchair directions $e_1$ and $e_2$.
This is because when $\mathbf{H}$ points along the zigzag direction $e_1$, the Grabner--Swanson equation gives \mbox{$\kappa_{ij}(H\hat{\mathbf{e}}_1) = \kappa'_{ij}\bigl(H(-\tfrac{1}{2}\hat{\mathbf{e}}_1+\tfrac{\sqrt{3}}{2}\hat{\mathbf{e}}_2)\bigr)$}, which is not a useful constraint because the field on the right side of the equation does not point along any of the three high-symmetry axes that we are interested in (i.e., \mbox{$e_1,e_2,e_3$}),
and similarly for when $\mathbf{H}$ points along the armchair direction $e_2$.
However, for the case where the magnetic field points perpendicular to the plane ($e_3$ direction), $C_3^{e_3}$ imposes the same constraints as for the field-independent case (see Eq.~\ref{eqn:rank-2_zero-field_tensor_constraints-C3e3_symmetry}), 
so \mbox{$\kappa_{ij}(H\hat{\mathbf{e}}_3)$} is of the form
\begin{equation}
\kappa_{ij}(H\hat{\mathbf{e}}_3) = 
\begin{pmatrix}
A & B & 0 \\
-B & A & 0 \\
0 & 0 & C
\end{pmatrix} \,. \qquad (C_3^{e_3})
\end{equation}
Identifying the symmetric and antisymmetric terms as even and odd functions of $H$, respectively, gives
\begin{equation}
\kappa_{ij}(H\hat{\mathbf{e}}_3) =
\overset{\overset{\textbf{\small Even in $H$}}{\vphantom{,}}}{\begin{pmatrix}
A & 0 & 0 \\
0 & A & 0 \\
0 & 0 & C
\end{pmatrix}}
+
\overset{\overset{\textbf{\small Odd in $H$}}{\vphantom{,}}}{\begin{pmatrix}
0 & B & 0 \\
-B & 0 & 0 \\
0 & 0 & 0
\end{pmatrix}} \,. \quad (C_3^{e_3}) \label{eqn:kappa_ij(He3)-constrained_by_C3e3}
\end{equation}

Finally, inversion symmetry does not impose any constraints on $\kappa_{ij}(\mathbf{H})$, since $\mathbf{H}$ is not affected by inversion, and $\mathbf{J}_Q$ and $\boldsymbol{\nabla} T$ both change sign under inversion, which leaves $\kappa_{ij}$ unchanged.

The most general forms of the magnetic-field-dependent thermal conductivity tensor $\kappa_{ij}(\mathbf{H})$ for the eight point groups generated by these three symmetries (see \mbox{Table~\ref{tab:point_groups}}) are given in \mbox{Table~\ref{tab:general_forms_of_tensors}}.
As an example, for a material of the monoclinic point group $2/m$ in an external magnetic field $\mathbf{H}$ along the zigzag axis $e_1$, the thermal Hall conductivity corresponding to a heat current $\mathbf{J}_Q$ along $e_1$ and a temperature gradient $\boldsymbol{\nabla} T$ along the armchair axis $e_2$ is given by the following boxed entry in \mbox{Table~\ref{tab:general_forms_of_tensors}}:
\begin{equation}
2,~2/m\hspace{-1.5pt}:~\kappa_{ij}(H\hat{\mathbf{e}}_1)
=
\overset{\overset{\textbf{\small Even in $H$}}{\vphantom{,}}}{\begin{pmatrix}
A & 0 & B \\
0 & C & 0 \\
B & 0 & D
\end{pmatrix}}
+
\overset{\overset{\textbf{\small Odd in $H$}}{\vphantom{,}}}{\begin{pmatrix}
\cline{2-2}
0 & \multicolumn{1}{|c|}{E} & 0 \\ \cline{2-2}
-E & 0 & F \\
0 & -F & 0
\\ \vspace{-13.6pt} \hphantom{-G_3} & \hphantom{-J_3} & \hphantom{-J_3}
\end{pmatrix} .}
\end{equation}

Similarly to the thermomagnetic susceptibility tensor discussed in Section~\ref{subsection:rank-3_zero-field_tensors}, even if components such as \mbox{$\kappa_{e_3 e_1}(\mathbf{H})$} and \mbox{$\kappa_{e_1 e_3}(\mathbf{H})$} are allowed by symmetry for a given point group and field orientation, we expect them to be zero for monolayer or few-layer systems.

\onecolumngrid
\begin{figure*}[ht!]
\begin{center}
\includegraphics[width=7.05in]{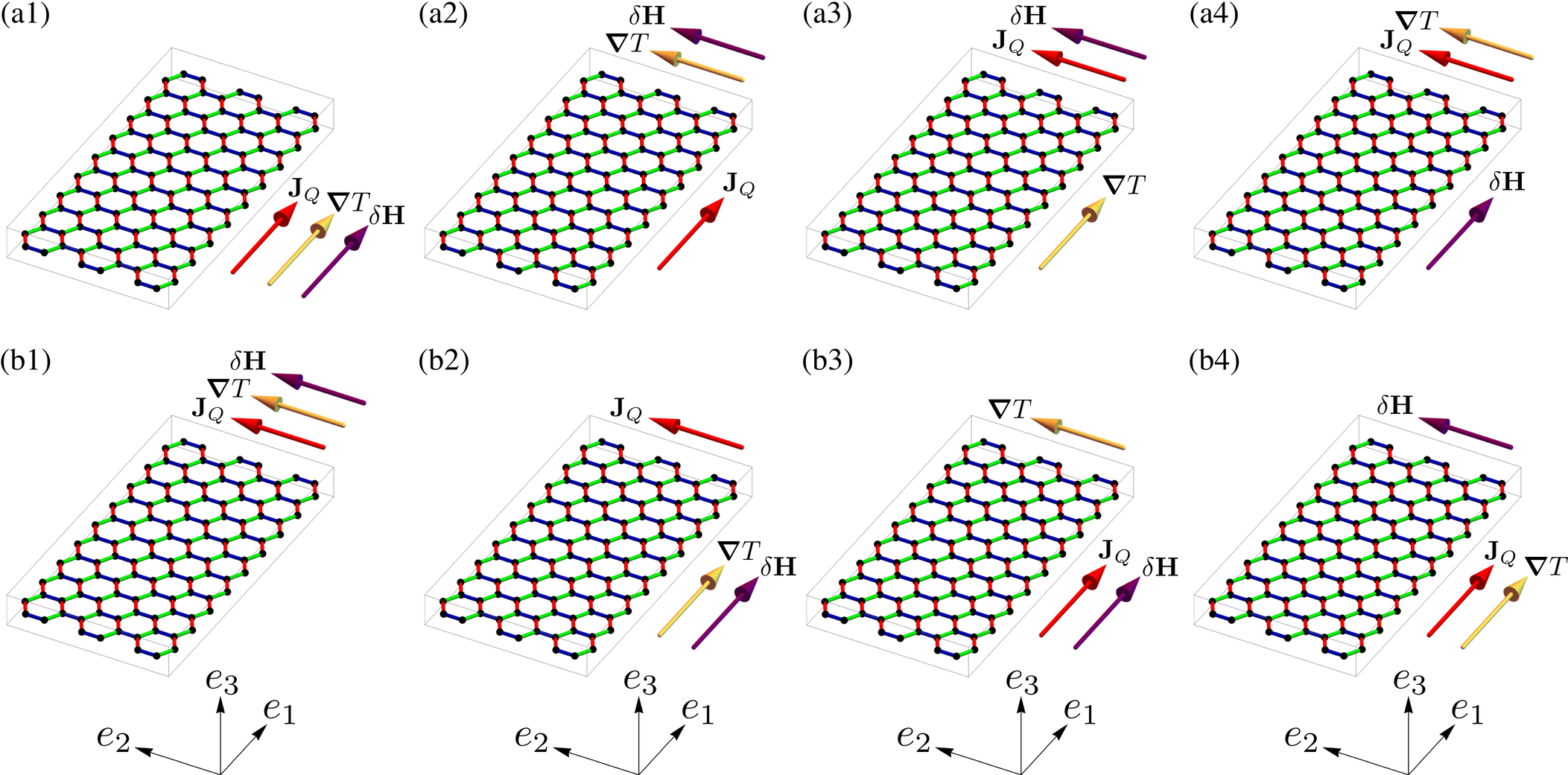}
\caption{Illustration of the two sets of unusual equalities (Eqs.~\ref{eqn:rank-3_zero-field_tensor_fully_longitudinal_equals_partly_transverse-top_line} and \ref{eqn:rank-3_zero-field_tensor_fully_longitudinal_equals_partly_transverse-bottom_line}) for zero-field \mbox{rank-3} tensors describing crystals with $C_3^{e_3}$ symmetry (i.e., belonging to the trigonal point groups $3$, $\bar{3}$, $32$, or $\bar{3}m$).
The changes in the thermal conductivity that results from applying a small magnetic field $\delta\mathbf{H}$ in scenarios \mbox{(a1)--(a4)} all have the same magnitude, i.e.
\mbox{$
|\delta\kappa_{e_1 e_1}(\delta\mathbf{H}\parallel\hat{\mathbf{e}}_1)|
=
|\delta\kappa_{e_1 e_2}(\delta\mathbf{H}\parallel\hat{\mathbf{e}}_2)|
=
|\delta\kappa_{e_2 e_1}(\delta\mathbf{H}\parallel\hat{\mathbf{e}}_2)|
=
|\delta\kappa_{e_2 e_2}(\delta\mathbf{H}\parallel\hat{\mathbf{e}}_1)|
$},
and similarly for scenarios \mbox{(b1)--(b4)}.
Reversing the direction of any one of these vectors simply reverses the sign of the change in the thermal conductivity.
These results hold when the system is either not magnetized, or magnetized along the out-of-plane direction ($e_3$).
Even though these illustrations describe the thermomagnetic susceptibility tensor \mbox{$\chi^\text{thermomag}_{ijk}=(\partial \kappa_{ij}/\partial H_k)|_{\mathbf{H}=\mathbf{0}}$} for concreteness, they more generally apply to any general \mbox{rank-3} tensor describing a system in the absence of external magnetic or electric fields by simply replacing the heat current $\mathbf{J}_Q$, temperature gradient $\boldsymbol{\nabla} T$, and small magnetic field $\delta\mathbf{H}$ with any three vector quantities as long as they are not large external fields (however, small magnetic or electric fields that are used to probe the system, i.e. $\delta\mathbf{H}$ and $\delta\mathbf{E}$, are allowed).}
\label{fig:thermomagnetic_susceptibility-all_predictions}
\end{center}
\end{figure*}
\twocolumngrid

\section{Summary of Predictions for Experiments}
\label{section:summary_of_predictions_for_experiments}
In this section we discuss the main predictions from our symmetry analysis and compare some of them with recent experiments.

\subsection{Predictions for Zero-Field Tensors}
We obtained the most general forms of \mbox{rank-2} and \mbox{rank-3} tensors expressed in the $e_1 e_2 e_3$ coordinates
\mbox{($e_1$ = zigzag direction, $e_2$ = armchair direction,} \mbox{$e_3$ = out-of-plane direction;} see Fig.~\ref{fig:honeycomb_lattice_of_edge-sharing_octahedra})
for crystals of various point groups in the absence of an external field. These results are listed in \mbox{Table~\ref{tab:general_forms_of_tensors}}.
We now highlight some notable predictions for these zero-field tensors for crystals of various point groups.

For systems with $C_3^{e_3}$ symmetry (i.e., crystals belonging to the trigonal point groups $3$, $\bar{3}$, $32$, or $\bar{3}m$ that are either not magnetized or magnetized along the out-of-plane direction) in the absence of an external field:
\begin{itemize}
    \item Rank-3 tensors have several unusual equalities between fully \mbox{\textit{longitudinal}} and partly \mbox{\textit{transverse}} in-plane components, as illustrated in Fig.~\ref{fig:thermomagnetic_susceptibility-all_predictions} using the thermomagnetic susceptibility tensor \mbox{$\chi^\text{thermomag}_{ijk}=(\partial \kappa_{ij}/\partial H_k)|_{\mathbf{H}=\mathbf{0}}$} (where $\kappa_{ij}$ is the thermal conductivity tensor) for concreteness.
    \item Rank-2 tensors have continuous rotational symmetry with respect to the axis perpendicular to the plane, so \mbox{rank-2} responses (e.g., magnetic susceptibility \mbox{$\chi_{ij} = (\partial M_i/\partial H_j)|_{\mathbf{H}=\mathbf{0}}$}) behave the same way along all in-plane directions, including low-symmetry directions.
\end{itemize}
For systems with $C_2^{e_2}$ symmetry (i.e., crystals belonging to the monoclinic point groups $2$ or $2/m$, or to the trigonal point groups $32$, or $\bar{3}m$ that are either not magnetized or magnetized along an axis that does not have $C_2$ symmetry) in the absence of an external field:
\begin{itemize}
    \item For tensors of all ranks, all tensor components corresponding to an odd number of \mbox{$e_1 + e_3$} directions are zero; for example, the following magnetic susceptibilities are zero: \mbox{$\chi_{e_1 e_2}$, $\chi_{e_2 e_1}$, $\chi_{e_3 e_2}$, $\chi_{e_2 e_3}$}.
\end{itemize}

\subsection{Predictions for the Thermal Conductivity Tensor \texorpdfstring{$\kappa_{ij}(\mathbf{H})$}{kappa ij(H)}}
We also obtained the most general forms of the thermal conductivity tensor $\kappa_{ij}(\mathbf{H})$ for crystals of various point groups in an external magnetic field along the high-symmetry directions $e_1$, $e_2$, and $e_3$ \mbox{($e_1$ = zigzag direction, $e_2$ = armchair direction,} \mbox{$e_3$ = out-of-plane direction;)}.
We looked at the components of $\kappa_{ij}(\mathbf{H})$ that are even and odd functions of the magnetic field $\mathbf{H}$ separately, where the even terms correspond to the thermomagnetic conductivity and the odd terms correspond to the thermal Hall conductivity.
These results are also listed in \mbox{Table~\ref{tab:general_forms_of_tensors}}.
We now highlight some notable predictions for the thermal conductivity tensor for crystals of various point groups.

For crystals with $C_3^{e_3}$ symmetry (i.e., belonging to the trigonal point groups $3$, $\bar{3}$, $32$, or $\bar{3}m$) in an external magnetic field:
\begin{itemize}
    \item When $\mathbf{H}$ points perpendicular to the plane (i.e., along $e_3$), $\kappa_{ij}(\mathbf{H})$ has continuous rotational symmetry with respect to this axis, so the thermal conductivity and thermal Hall conductivity behave the same way along all in-plane directions, including low-symmetry directions.
\end{itemize}
For crystals with $C_2^{e_2}$ symmetry (i.e., belonging to the monoclinic point groups $2$ or $2/m$, or to the trigonal point groups $32$, or $\bar{3}m$) in an external magnetic field:
\begin{itemize}
    \item In an external magnetic field $\mathbf{H}$ along the in-plane zigzag axis $e_1$, applying a heat current $\mathbf{J}_Q$ along $e_1$ can produce a thermal Hall response (i.e., a transverse temperature gradient $\boldsymbol{\nabla} T$ along the high-symmetry armchair axis $e_2$ that reverses direction upon reversing the direction of $\mathbf{H}$). This has been observed in a recent thermal Hall experiment on \mbox{$\alpha$-RuCl$_3$} (belonging to the monoclinic point group $2/m$~\cite{Johnson2015}) by Yokoi et al.~\cite{Yokoi2020} (see Fig.~\ref{fig:thermomagnetic_susceptibility-one_example_of_fully_longitudinal_equals_partly_transverse}a with $\delta\mathbf{H}$ replaced by $\mathbf{H}$ for an illustration of the orientations used in this experiment),
    as well as corroborated analytically and numerically by Chern, Zhang, \& Kim~\cite{Chern2020,Zhang2021}.
    \item When $\mathbf{H}$ is along the in-plane high-symmetry armchair axis $e_2$, applying a heat current $\mathbf{J}_Q$ along the zigzag axis also $e_1$ \textit{cannot} produce a thermal Hall response (i.e., a transverse temperature gradient $\boldsymbol{\nabla} T$ along $e_2$ that reverses direction upon reversing the direction of $\mathbf{H}$). This was also observed in \mbox{$\alpha$-RuCl$_3$} by Yokoi et al.~\cite{Yokoi2020} and corroborated analytically and numerically by Chern et al.~\cite{Chern2020,Zhang2021}.
    \item When $\mathbf{H}$ is along the in-plane high-symmetry armchair axis $e_2$, applying a heat current $\mathbf{J}_Q$ along $e_2$ \textit{cannot} produce a thermal Hall response (i.e., a transverse temperature gradient $\boldsymbol{\nabla} T$ along the zigzag axis $e_1$ that reverses direction upon reversing the direction of $\mathbf{H}$). Relative to the orientations described in the first bullet point, this corresponds to interchanging which vectors point along $e_1$ and $e_2$ (i.e., \mbox{$e_1 \leftrightarrow e_2$}), or equivalently, to rotating the three vectors by $90^\circ$ with respect to the out-of-plane axis.
\end{itemize}

Other experiments have also observed a thermal Hall effect in \mbox{$\alpha$-RuCl$_3$} when the magnetic field is applied in the plane, although the direction of the field within the plane was not known~\cite{Hentrich2018,Hentrich2019,Hentrich2020}.
More experiments are needed to get a better understanding of the tensorial character of the thermal Hall response in these materials.

\section{Outlook}
\label{section:outlook}
This work has the potential to guide future experiments seeking to probe new physical responses along different geometries in 2D materials, similarly to the unusual thermal Hall effect observed in \mbox{$\alpha$-RuCl$_3$} when the magnetic field is applied in the plane~\cite{Yokoi2020}.
Our analysis can also help inform the search for existing 2D materials or the design of novel materials having specific desirable properties (e.g., the presence or absence of a given longitudinal or transverse physical response).
Finally, this analysis can aid in the identification of the crystal structure (specifically, the point group) of new 2D materials.
The analysis presented here can also be extended to the magnetic point group symmetries following a similar procedure.

\section{Acknowledgments}
We thank Joshua E. Goldberger for useful discussions about crystal symmetries and point groups in these 2D systems. We also thank Joseph P. Heremans and Rolando Valdés Aguilar for their helpful feedback and for pointing us to various relevant references. Finally, we thank Brian Skinner, Zachariah Addison, Joseph Szabo, Humberto Gilmer, and Daniella Roberts for their feedback.
This research was partially supported by the Center of Emergent Materials, an NSF MRSEC under award number DMR-2011876, and from BES-DOE grant DE-FG02-07ER46423.

\bibliography{main}

\end{document}